\documentclass[a4paper,11pt]{article} 

\usepackage{jinstpub}
\usepackage{lineno}
\usepackage{scrextend}

\usepackage[lofdepth=1]{subfig}
\usepackage{float}
\usepackage{gensymb}
\usepackage{upgreek}
\usepackage{gensymb} 
\usepackage[section]{placeins}

\title{\boldmath First stave results towards mitigating sensor fracturing with interposers in the ATLAS ITk strips barrel}

\author[b]{G. D'Amen,}
\author[d]{D. Dewhurst,}
\author[d]{E. Dibley,}
\author[d]{J. Dopke,}
\author[a]{E. Duden,}
\author[d]{G. Hawker,}
\author[d]{\\B. Gallop,}
\author[d]{N. Ghorbanian,}
\author[c]{P. Jacobson,}
\author[b]{M. Kurth,}
\author[b]{A. Li,}
\author[b]{D. Lynn,}
\author[c]{A. Petersen,}
\author[d]{\\P. Phillips,}
\author[d]{D. Russell,}
\author[d]{C. Sawyer,}
\author[e]{C. Solaz,}
\author[a]{W. Sorger,}
\author[b]{S. Stucci,}
\author[b,*]{\\A. Tishelman-Charny, \note[*]{Corresponding author.}}
\author[b]{A. Tricoli,} 
\author[b]{and G. van Nieuwenhuizen}

\affiliation[a]{Department of Physics, Brandeis University, Waltham MA}
\affiliation[b]{Brookhaven National Laboratory (BNL), Upton, NY 11973 (NY), U.S.A}
\affiliation[c]{Department of Physics, Duke University, Durham NC, U.S.A}
\affiliation[d]{Particle Physics Department, STFC Rutherford Appleton Laboratory, Harwell Science and Innovation Campus, Didcot, United Kingdom}
\affiliation[e]{Instituto de Física Corpuscular (IFIC) - CSIC-University of Valencia, Parque Científico, C/Catedrático José Beltrán 2, E-46980 Paterna, Spain}

\emailAdd{abraham.tishelman.charny@cern.ch}

\abstract{At the conclusion of Run 3 of the Large Hadron Collider (LHC) at CERN, the accelerator complex will be upgraded to the High-Luminosity LHC (HL-LHC), allowing it to increase the dataset sizes of LHC experiments by about a factor of 20. This significant increase in dataset size will improve the sensitivity and precision of all physics analyses from LHC experiments, but will come with a more challenging data-taking environment. In order to handle this, the ATLAS detector will undergo a substantial upgrade, including an upgrade of its inner tracker to an all-silicon tracker called the Inner Tracker (ITk), made of pixel and strip sub-detectors. During the pre-production phase of the ITk strips, it was discovered that thermally cycling modules loaded onto local support structures led to physical fractures in silicon sensors due to the induced thermal stress. This is understood to be the result of several factors, including the difference in coefficients of thermal expansion between the different module layers, and the close proximity of the module electrical components. Several mitigation strategies were tested to reduce the rate of module fracturing. This paper describes the assembly setup, testing setups, and electrical testing results of ITk strips barrel modules loaded onto local support structures. 69 of 70 modules with an in-built additional kapton layer were found to survive testing down to -70$\degree$C.}


\begin{document}

\maketitle
\flushbottom

\section{Introduction}
\label{sec:introduction}

At the end of Run 3 of the Large Hadron Collider (LHC), the accelerator complex will be upgraded in order to increase the datasets of its experiments by a factor of about 20. To cope with a more challenging data-taking environment, the ATLAS~\cite{ATLAS} detector will undergo major upgrades, including an upgrade of its tracker to an all-silicon tracker called the ITk, composed of pixel and strip sub-detectors~\cite{ITkTDR}, where the strip sub-detector is composed of a barrel and two endcaps. The ITk strip detector will be composed of modules, assembled by gluing Printed Circuit Board (PCB) flexes directly onto silicon sensors. Modules are then glued directly onto local support structures, which will be inserted into global mechanical structures.

During the pre-production phase of construction of the ITk strip detector, it was discovered that thermally cycling modules loaded onto local support structures formed physical fractures in modules. This is understood to primarily be the result of a Coefficient of Thermal Expansion (CTE) mismatch between different layers of the module, creating regions with high strain when performing tests at -35$\degree$C \cite{simulation}. As this is the planned operational temperature of the ITk, a robust mitigation solution is necessary to ensure modules are able to operate cold.

Three mitigation strategies were explored: A stiffer loading glue for attaching modules to local support structures, an increased distance between module components, and the addition of an extra layer of glue and polyimide (Kapton) between the sensor and electronic components. All three mitigation strategies were tested with different modules loaded onto barrel local support structures, called staves.

This paper is structured as follows: section \ref{sec:modules_and_staves} describes modules and staves. Section \ref{sec:Stave_Assembly} describes an example stave assembly setup. Section \ref{sec:Stave_Testing} describes the two stave testing setups and testing procedure performed at the two stave testing sites. Section \ref{sec:Results} describes the electrical testing results of mitigation staves, and section \ref{sec:conclusions} describes the conclusions of the paper.
\section{Modules and Staves}
\label{sec:modules_and_staves}

\begin{figure}[htbp]
\centering
\includegraphics[width=\textwidth]{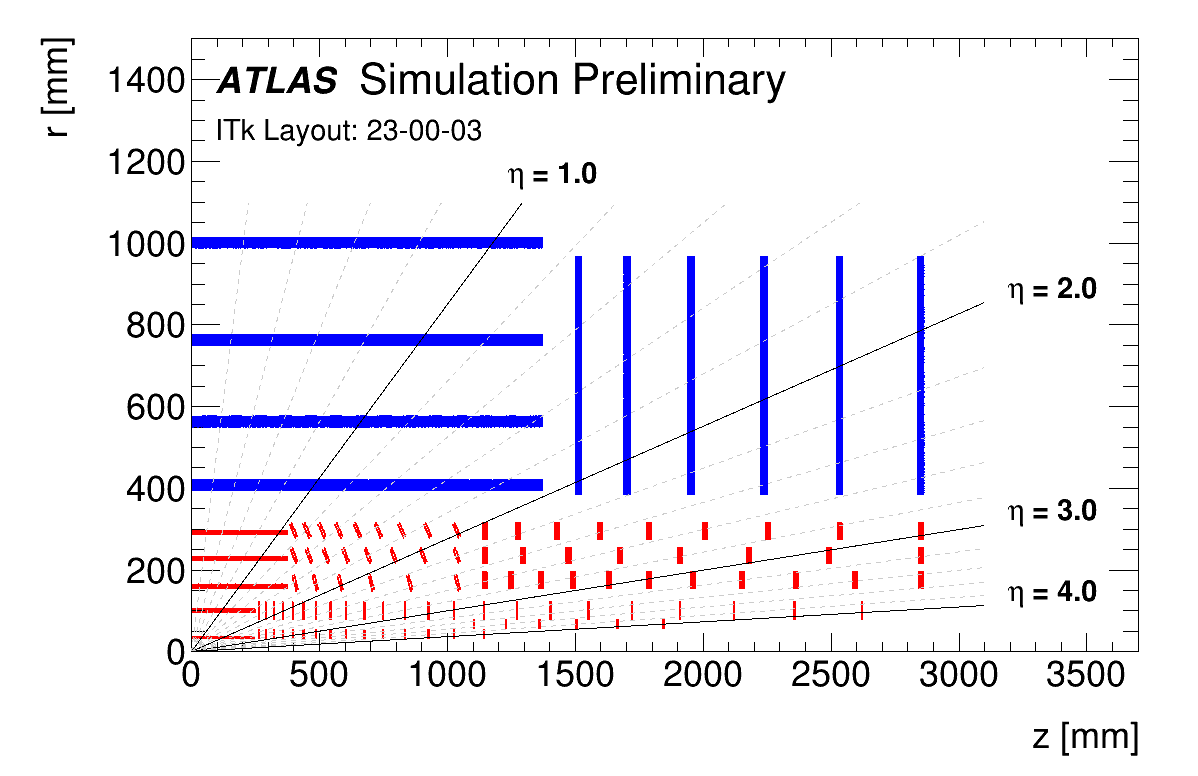}
\caption{ 
The ITk layout~\cite{ITk_Layout}. The four barrel layers correspond to the four horizontal blue lines ranging from about r = 400~mm to r = 1000~mm.
\label{fig:ITk_layout}}
\end{figure}

The layout of the full ITk is shown in Figure \ref{fig:ITk_layout}. The ITk strips barrel will be made of four concentric layers, called layers L0-L3. Layers L0 and L1 will be closer to HL-LHC particle collisions, and will be made of Short Strip (SS) staves. Layers L2 and L3 will be further from collisions, and will be made from Long Strip (LS) staves. A total of 256 LS staves will be made to make up the L2 and L3 layers of the Barrel.

\begin{figure}[htbp]
\centering
\includegraphics[width=\textwidth]{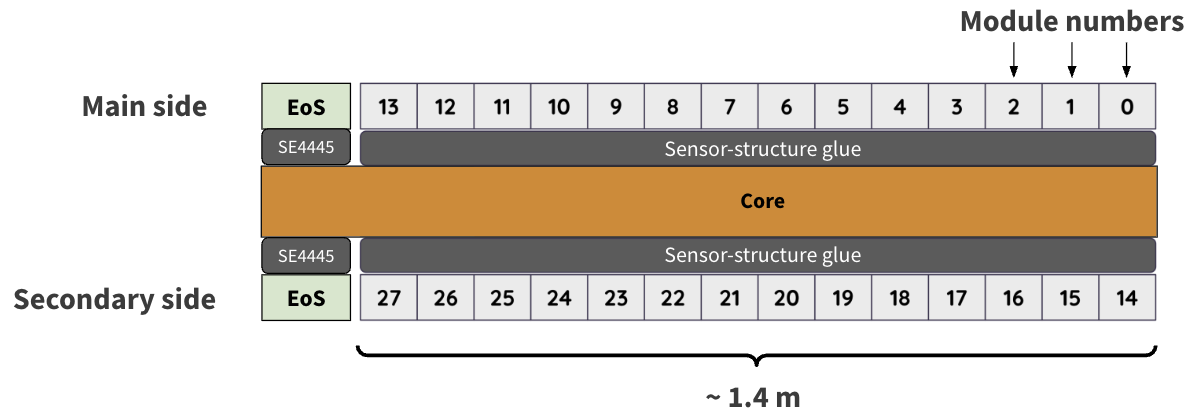}
\caption{ 
Illustrative diagram of a stave from the side perspective (not to scale) showing the ``main" and ``secondary" sides, each containing 14 modules. An EoS board is present on each side of the core, and modules are glued to the core.
\label{fig:stave_diagram}}
\end{figure}

A stave is assembled from a local support structure called a ``core", large poliymide flex circuits attached to the surfaces called ``bustapes", PCBs called End of Structure boards (EoS) \cite{EOS} that provide connection to the rest of the detector, attached to the core with Dow Corning SE 4445 CV (``SE4445") \cite{SE4445}, and standalone silicon strips modules. The EoS hosts lpGBTx chips for data readout~\cite{lpgbt}, DCDC converters \cite{BPOL}, and a VTRx \cite{VTRx} for digital to optical conversion. Included in the core is a loop of titanium piping, with an inlet and outlet, allowing coolant to flow through the core and make a loop back through the outlet. A diagram of a stave is shown in Figure \ref{fig:stave_diagram}. Further details of the formation of a stave before the addition of modules can be found in \cite{LocalSupports}.

\begin{figure}[htbp]%
    \setcounter{subfigure}{0}
    \centering
    \subfloat[An example barrel module diagram, for an SS module. Reproduced from \cite{TDR}. CC BY 4.0.\label{fig:module_diagram}]{\includegraphics[height=0.25\textheight]{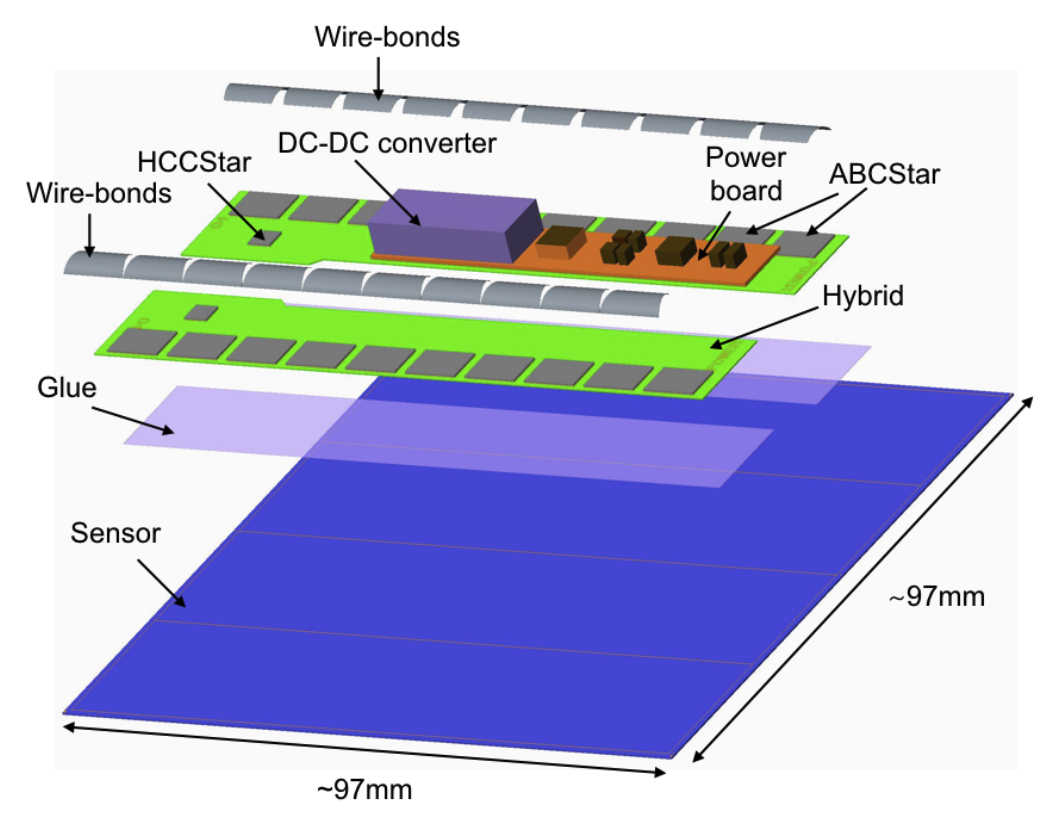}}%
    \hfill
    \subfloat[A photo of a Long Strip module, with streams 0 and 1 labeled. Note that strips are aligned in the vertical direction. Reproduced from \cite{ModuleImage}.~\textcopyright~2022 IOP Publishing Ltd and Sissa Medialab. All rights reserved.]{\includegraphics[height=0.25\textheight]{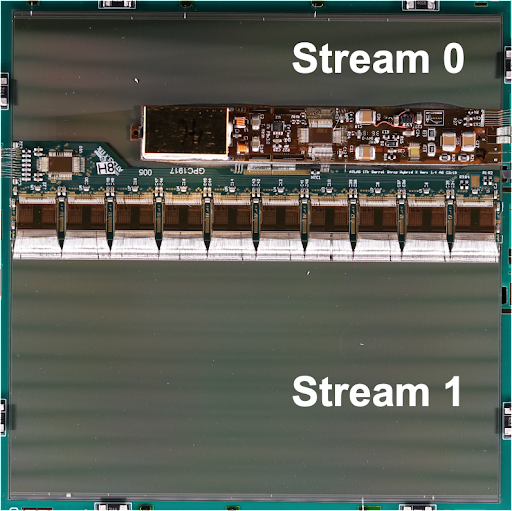}}%
    \caption{Diagram and picture of a module.\label{fig:module_image}}%
\end{figure}

The individual detection unit of the ITk strips is a module. A module is made by starting with a silicon strip sensor, and gluing PCBs on top which host the module's data readout and powering circuits. Barrel modules come in two types: Long strip and short strip. Both are about 10 cm by 10 cm in size, but SS modules have twice the number of strips. As the plan for Barrel module production is to begin with LS modules and later transition to SS modules, this paper will focus on LS modules. A diagram and picture of a module is shown in Figure \ref{fig:module_image}.

An LS module is partitioned into two rows, each containing 1280 strips, referred to as Streams 0 and 1. A single flexible PCB, called a hybrid, is glued onto the sensor, and hosts 11 ASICs glued to its surface: 10 ABCStars \cite{ABCStar} used for DAQ readout, and one HCCStar \cite{HCCStar} to control the ABCStars. The hybrid is equipped with an NTC thermistor for temperature readout, and the individual strips are wirebonded to the ABCstars for readout.

A second flexible PCB, the powerboard, is also glued to the sensor. The powerboard is equipped with a central ASIC called the AMACstar \cite{AMACStar}, used to monitor module quantities and send single ended signals to both the HCCStar and ABCStars.


During the pre-production phase of ITk strips module assembly, it was discovered that some modules loaded onto staves would form physical fractures when cooled to -35$\degree$C. This is understood to be primarily due to a CTE mismatch: Hybrids and powerboards are composed of copper and polyimide, which both have much larger CTEs than silicon. This means when modules are cooled to low temperatures, hybrids and powerboards tend to contract much more than sensors. Furthermore, because of the close proximity of the hybrid and powerboard on the module, and because the gluing of the sensor onto the core prevents the sensor from moving with the hybrid and powerboard, strain is induced with an expected maximum value located between the hybrid and powerboard \cite{simulation}. In order to find a solution to mitigate this effect, three strategies were tested: The use of a stiffer loading glue between modules and staves, changing from the nominal adhesive from SE4445 to Loctite EA 9396 Aero (``Hysol") \cite{Hysol}, an increased distance between the powerboard and hybrid from 1mm to 3mm (``wide-gap"), and the addition of an extra layer of glue and polyimide (Kapton) between the sensor and electronic components (``Interposer"). Further explanations of the sensor fracturing mechanism, signature in electrical tests, and different mitigation strategies can be found in \cite{LHCP_Proceedings}.

The details of the previously standard non-interposed module assembly procedure can be found in \cite{ITkstripsQC}, and the details of the new interposer module assembly procedure can be found in \cite{Anne_Proceedings}.
\section{Stave assembly}
\label{sec:Stave_Assembly}

There are two institutes that assemble staves, henceforth referred to as site 1 and site 2. Each stave assembly site receives the following in order to assemble a stave, as previously described in Section \ref{sec:modules_and_staves}: assembled cores, modules, and EoS boards. Each stave assembly site is in charge of gluing the EoS boards and the modules onto the cores, making the electrical connections of the modules and the EoS boards to the bustape pads by means of wire-bonding, and performing the electrical testing and quality control of the completed staves. The most challenging part of this assembly procedure is guaranteeing the XY precision of the silicon modules, which must be placed to within 50~$\upmu$m in XY. The two sites have very similar stave assembly setups - therefore, this section will only describe stave assembly at site 1. So far, a total of 8 staves have been assembled in order to test different sensor fracturing mitigation strategies, as summarized in Table \ref{fig:Stave_Summary}.

\begin{table}[htbp]
    \centering
    \includegraphics[width=\linewidth]{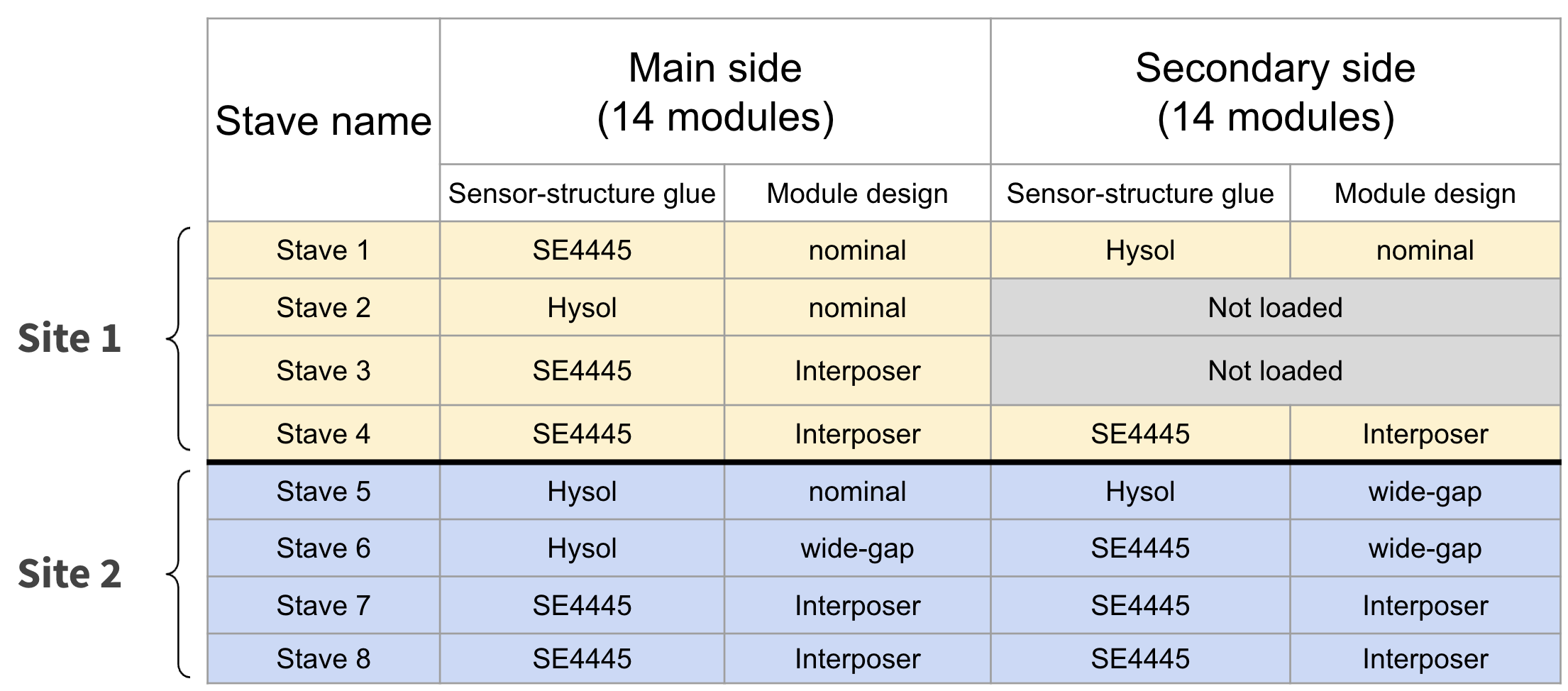}
    \caption{Summary of fracturing mitigation staves. Staves covering all three major testing mitigation strategies were assembled: Using Hysol for loading instead of SE4445, the wide-gap module design, and interposers.}
    \label{fig:Stave_Summary}
\end{table}

\begin{figure}[!htb]
    \centering
    \includegraphics[width=\linewidth]{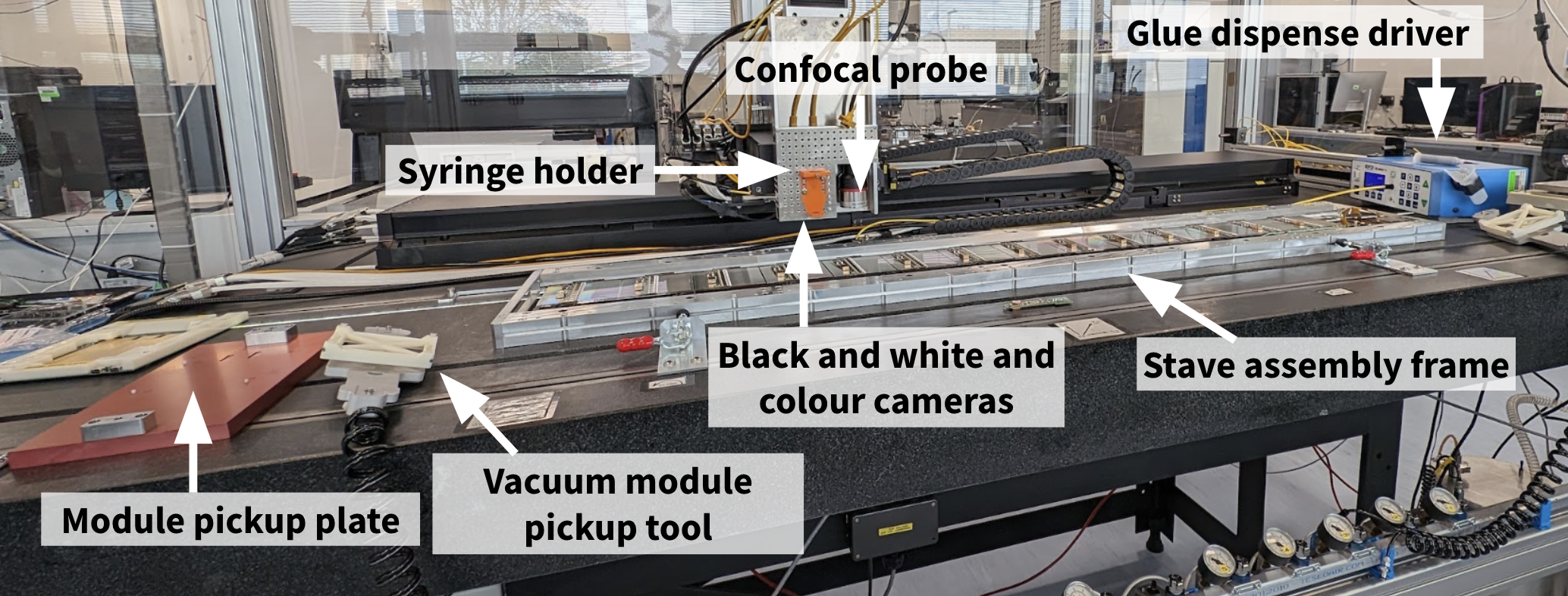}
    \caption{An overview of the module loading (the placement of modules onto the stave) setup for site 1. Glue is accurately dispensed onto the core, while the module sits on the pickup plate before being picked up with the vacuum module pickup tool, and is then placed onto the core.}
    \label{fig:Site_1_GantryOverview}
\end{figure}

Stave assembly is performed on a granite table with a robotic 3-axis gantry in charge of the glue dispensing, metrology, and visual capture. For stave assembly at site 1, the stave core is placed in an assembly frame on a granite gantry table as shown in Figure \ref{fig:Site_1_GantryOverview}. Additionally, an EFD Ultimus V pressure glue dispenser \cite{EFD} is situated to the side of the assembly area, and its syringe is mounted on the gantry. The movements are controlled by software written in Matlab \cite{matlab}. Next to the glue dispenser on the gantry, there is a black and white camera, a color camera, and a Keyence CL-P070 confocal laser \cite{keyence}, that are controlled using the same instance of Matlab software. All of this is necessary in order to achieve the necessary XY precision of the silicon modules.

A calibration procedure is performed in order to align the glue dispensing system with the core. A dedicated software is used to define a coordinate system with respect to lockpoints on the core - further details can be found in Appendix \ref{appendix:AssemblyDetails}. After calibration is completed, module loading can begin. The glue dispenser mounted on the gantry is used to dispense glue for the EoS boards and modules. The software allows for a finely controlled glue pattern to be programmed into the dispenser. 

\begin{figure}[!htb]
    \centering
    \includegraphics[width=0.5\linewidth]{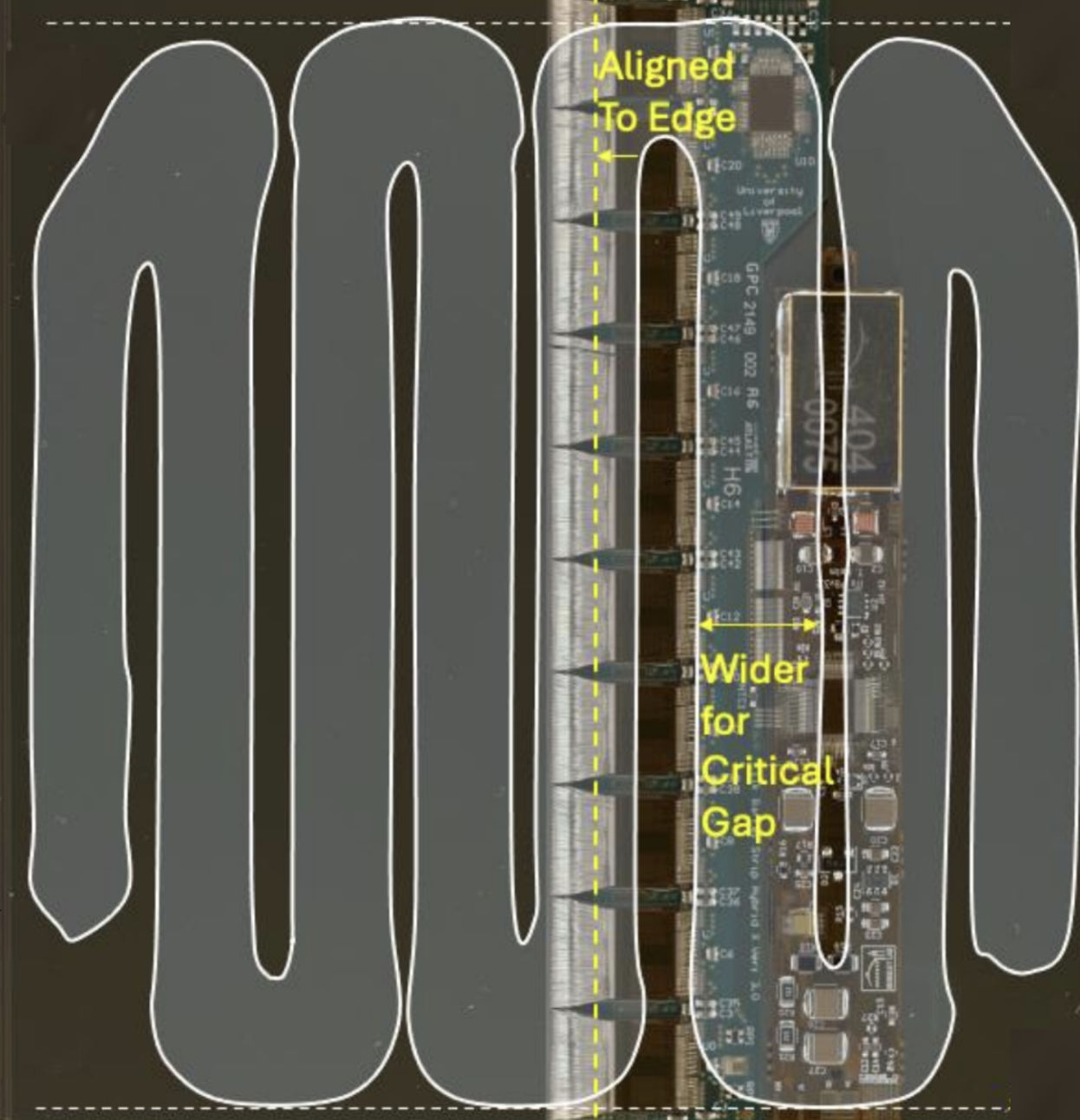}
    \caption{The updated module to core glue pattern used for LS modules, where a wider line is used for the critical gap between the hybrid and powerboard.}
    \label{fig:SnakePattern}
\end{figure}

\begin{figure}[!htb]
    \centering
    \includegraphics[width=\linewidth]{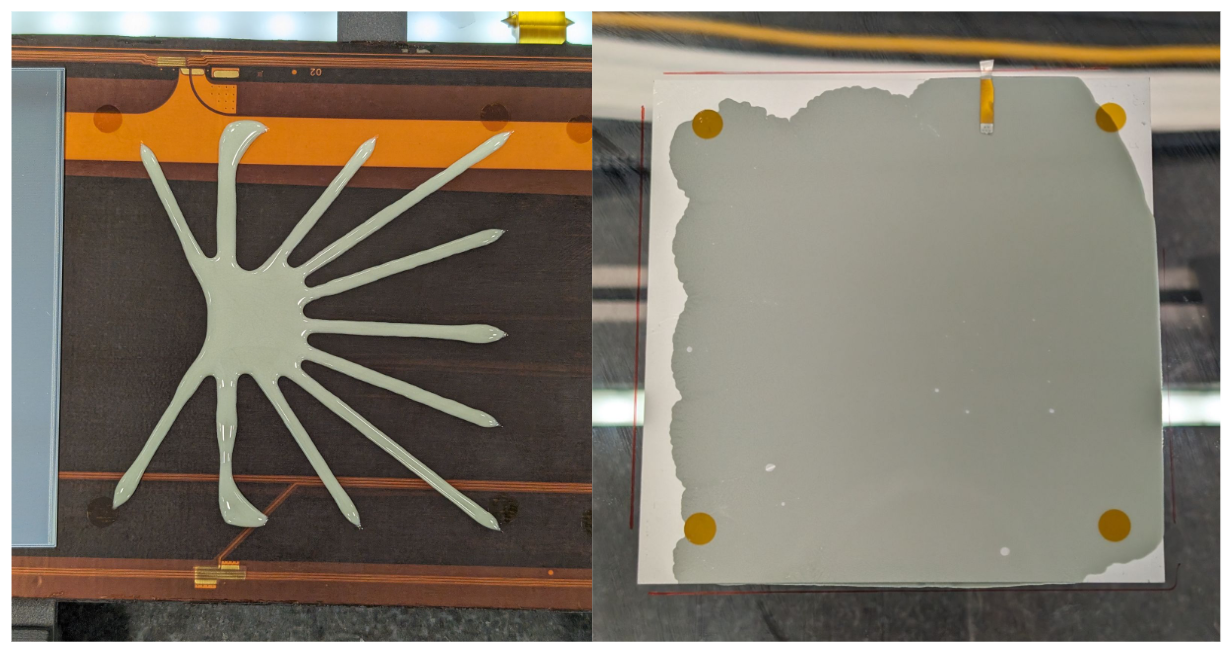}
    \caption{The stave 2 glue pattern, which differed from all other staves, aimed at full coverage under the module. Left: the glue pattern as dispensed. Right: the glue pattern under compression of a long strip module mounted on a glass slide.}
    \label{fig:Site_1_FullCoverageHysol}
\end{figure}

A part of the modifications of the default procedures included adjusting the module to core glue patterns. For LS modules, a ``snake" pattern is used as the nominal pattern, with a thicker glue line placed to sit under the gap between the powerboard and hybrid, shown in Figure \ref{fig:SnakePattern}. For all mitigation staves except one, this modification is included to improve the coverage under the area where maximum stress is expected. One stave from site 1, which used Hysol as glue, incorporated a completely new glue pattern aiming for full coverage under the sensor, as shown in Figure \ref{fig:Site_1_FullCoverageHysol}.

Glue is dispensed for a single module at a time in this setup. The module is picked up using a vacuum pickup tool, which holds the module with suction cups at its four corners. Using this technique, modules can be positioned within~$\pm50\upmu$m in X and Y. A screenshot of the locating of the fiducials in the loading software is shown in Figure \ref{fig:Site_1_Fiducials}. Further details on module placement can be found in Appendix \ref{appendix:AssemblyDetails}.

\begin{figure}[!htb]
    \centering
    \includegraphics[width=\linewidth]{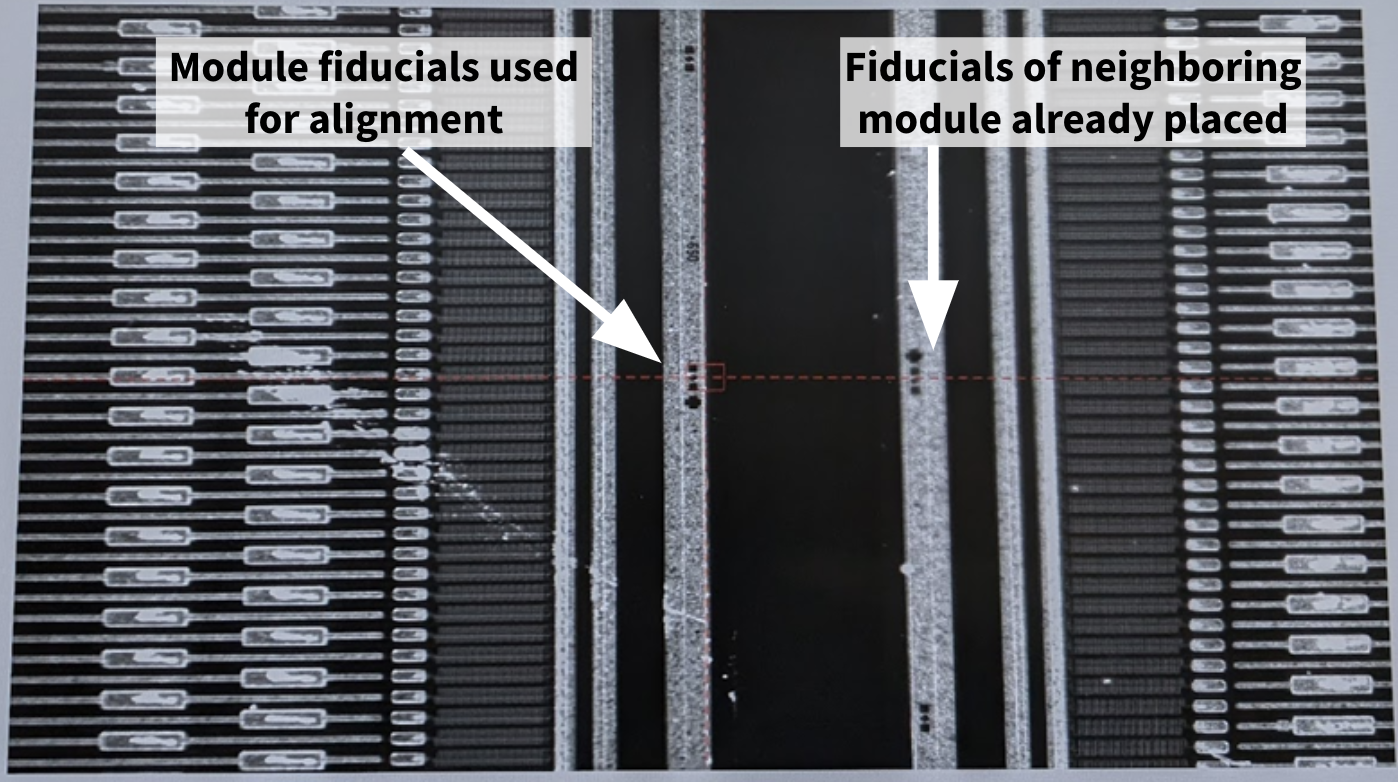}
    \caption{The fiducials of a module during loading, and its neighboring module, on a display screen used during fine adjustments of module position.}
    \label{fig:Site_1_Fiducials}
\end{figure}
\section{Stave testing}
\label{sec:Stave_Testing}

The final step in the quality control (QC) of the finished staves consists of performing electrical tests on the modules using the built-in test circuitry equipped in the readout ASICs. Through these tests, input noise can be measured and problematic channels (classified as having unusual response patterns) can be identified. Current-voltage curves (IV) are obtained to make sure the sensors do not present early breakdowns at the operating voltages, indicated by a sudden slope change greater than a factor of 5, and a leakage current greater than $\approx$10~$\upmu$A. Noise measurements are also taken to ensure relatively uniform noise across strips. Examples of fracture signatures in electrical tests due to fracturing are shown in Figure 3 of \cite{LHCP_Proceedings}. Common fracturing signatures are quasi-linear current changes with a large slope in IVs, and non-uniform noise among localized groups of strips in noise measurements.

As the operational temperature of the ITk will be -35$\degree$C, it must be ensured that modules loaded onto staves can still produce quality data at this temperature. For this reason, as part of nominal stave QC, electrical tests are to be taken at room temperature and -35$\degree$C. However, in the stages of design validation, temperatures below -35$\degree$C have been used to assess the headroom of different module mitigation designs. The different testing temperatures for design validation are described further in Section \ref{sec:testingProcedure}.

For testing the staves, Faraday boxes with nitrogen or dry air inlets to guarantee low relative humidity (RH) operation, electrical connections for operating the staves, and connection to the cooling tubes are required.

Both stave assembly sites are equipped with stave testing setups for electrical testing. This section will describe the sites' stave testing setups, and the procedure for electrically testing staves.

\subsection{Site 1 \label{Testing_Site_1}}

\begin{figure}[htbp]
    \centering
    \includegraphics[width=0.6\linewidth]{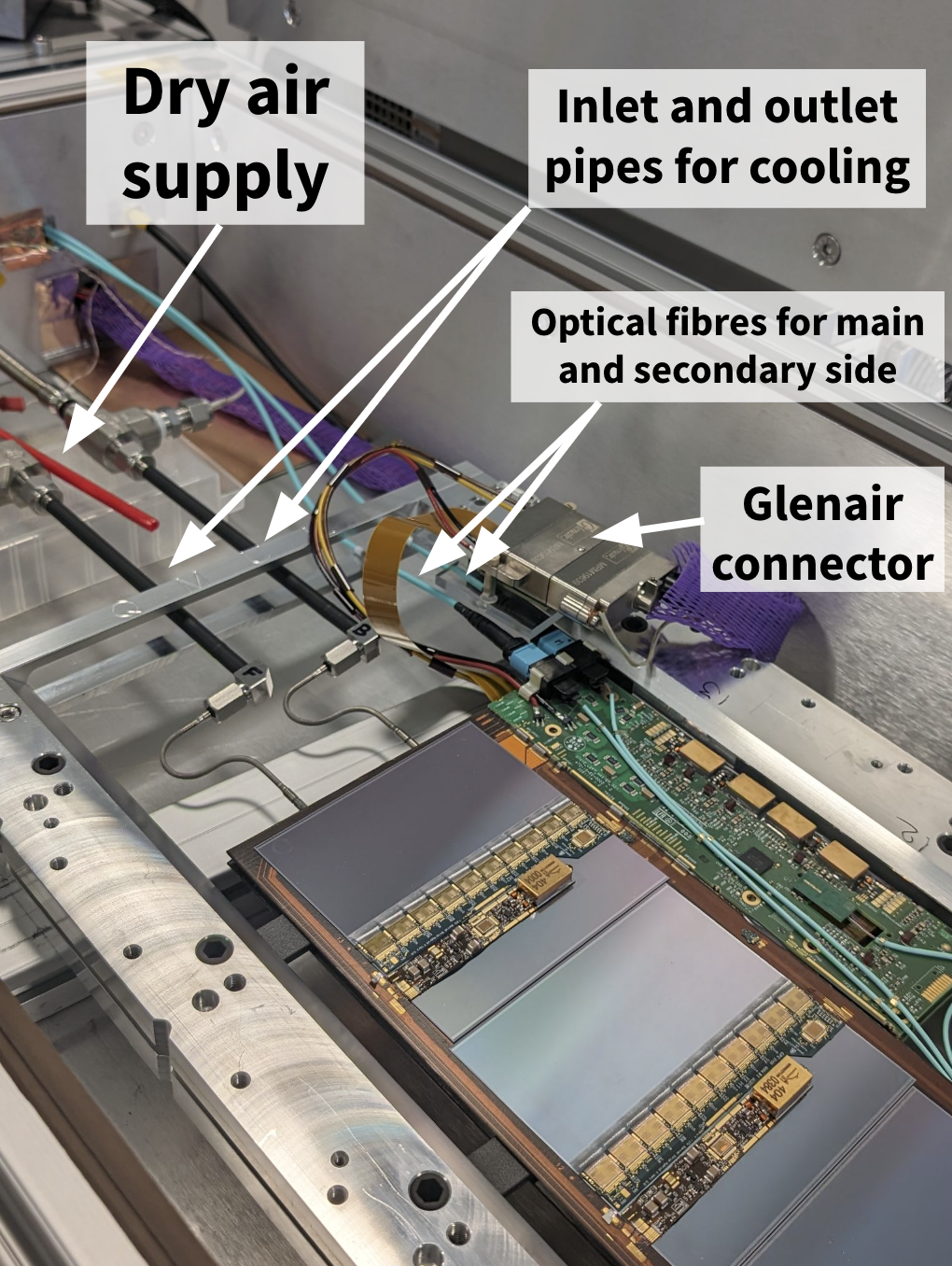}
    \caption{The stave box at site 1 for testing staves down to -45$\degree$C. The following are pictured and labeled: A ``dry air flow" tube, which receives dry air from the wall and is directed to the stave to ensure moisture is not present on the modules, a crucial precaution as the sensors are sensitive to moisture. A custom connector (Glenair connector) for providing low voltage, high voltage and NTC readings, an EoS board, and optical fibers used for communication with the Genesys evaluation board of the Kintex-7 FPGA. These optical fibers will also be used for off-stave communication when staves are installed in the detector.}
    \label{fig:Setup_1_StaveTestBox}
\end{figure}

\begin{figure}[htbp]
    \centering
    \includegraphics[width=0.7\linewidth]{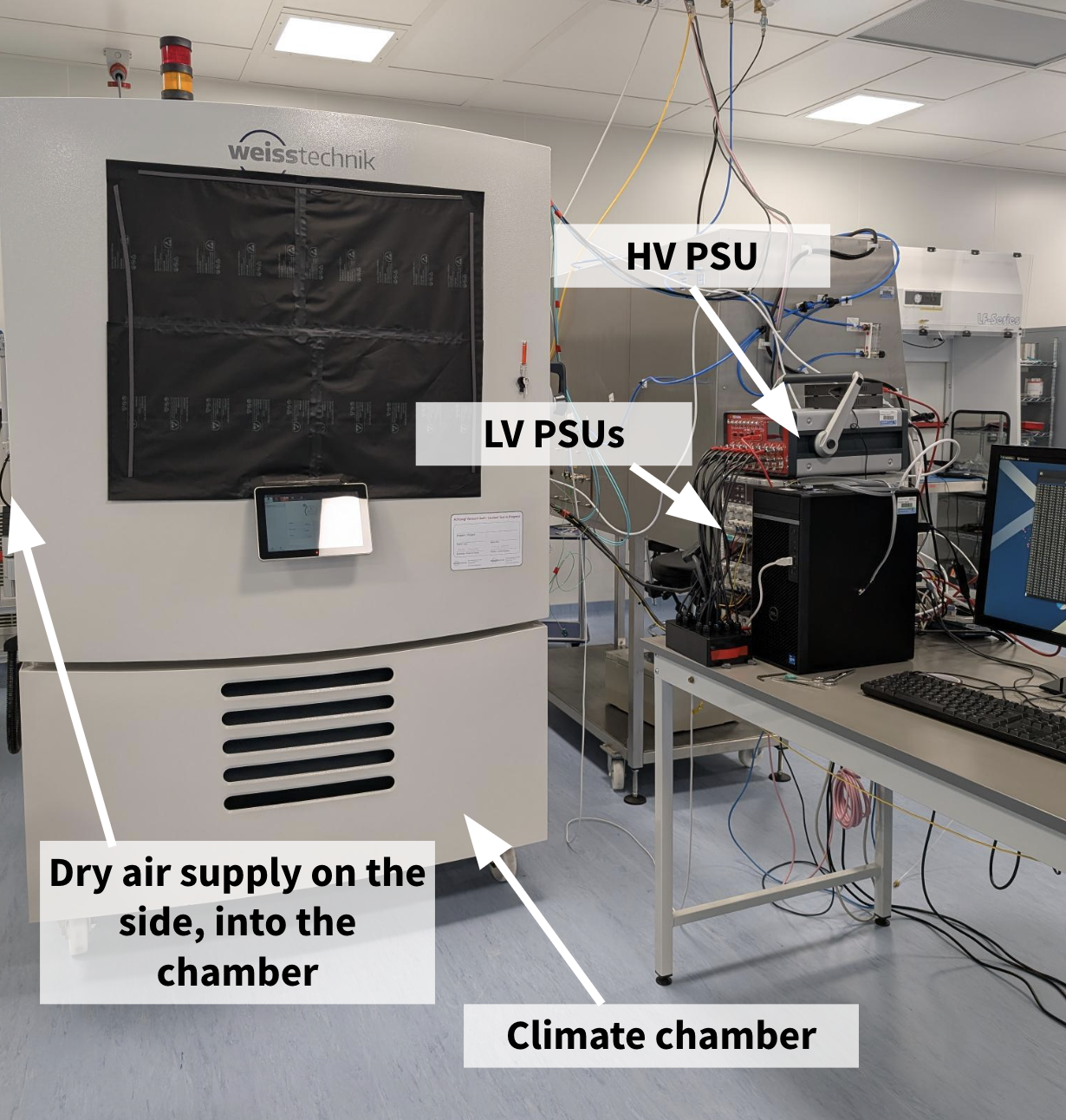}
    \caption{The climate chamber setup at site 1 for testing staves down to -70$\degree$C. Staves are placed inside the climate chamber, which has dry air flowing through to minimize humidity on the sensors. Low Voltage (LV) and High Voltage (HV) Power Supply Units (PSUs) are used to power the EoS boards, and bias the sensors respectively.}
    \label{fig:Setup_1_CC}
\end{figure}

Site 1 has two stave testing setups: a stave test box, very similar to what will be described in Section \ref{Testing_Site_2}, and a climate chamber for testing at colder temperatures. The stave test box is shown in Figure \ref{fig:Setup_1_StaveTestBox}, and can reach a minimum average stave inlet/outlet temperature of -42$\degree$C. The climate chamber is shown in Figure \ref{fig:Setup_1_CC}, and uses air cooling that can reach -70$\degree$C. Included in the setup are Low Voltage (LV) and High Voltage (HV) Power Supply Units (PSUs).

\begin{figure}[htbp]
    \centering
    \includegraphics[width=0.75\linewidth]{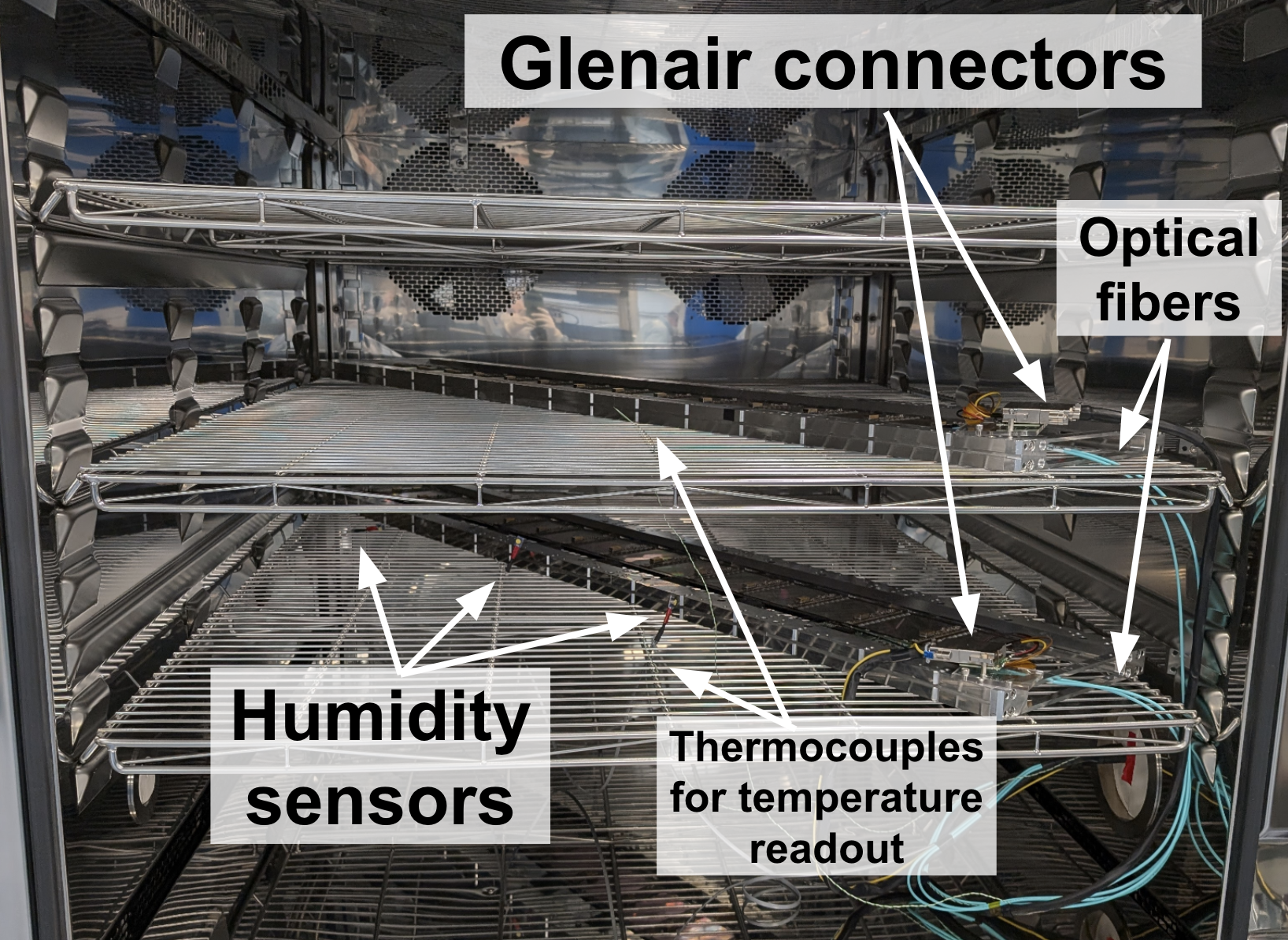}
    \caption{Two long strip staves in the climate chamber at setup 1. Several humidity sensors are placed in order to ensure very low humidity, to avoid condensation on modules. Thermocouples are also placed to monitor air temperature as a proxy for sensor temperature. Glenair connectors are present to provide LV and HV from the PSUs to the stave, and optical fibers are present to transfer data after conversion from digital format.}
    \label{fig:Setup_1_CC_Stave_Connections}
\end{figure}

The climate chamber is configured to be able to cool and read out two long strip staves simultaneously. The connections to the staves and the humidity and temperature sensors inside the chamber are shown and labeled in Figure \ref{fig:Setup_1_CC_Stave_Connections}. The temperature sensors are K-type sensors \cite{Ktype_NTC} that can readout down to -40$\degree$C.

Similar to the stave test boxes, the climate chamber stave testing setup is equipped with a software monitoring system, allowing for the constant monitoring and logging of environmental and DAQ (Data Acquisition) related quantities. 


\subsection{Site 2 \label{Testing_Site_2}}

\begin{figure}[htbp]
    \centering
    \includegraphics[width=\linewidth]{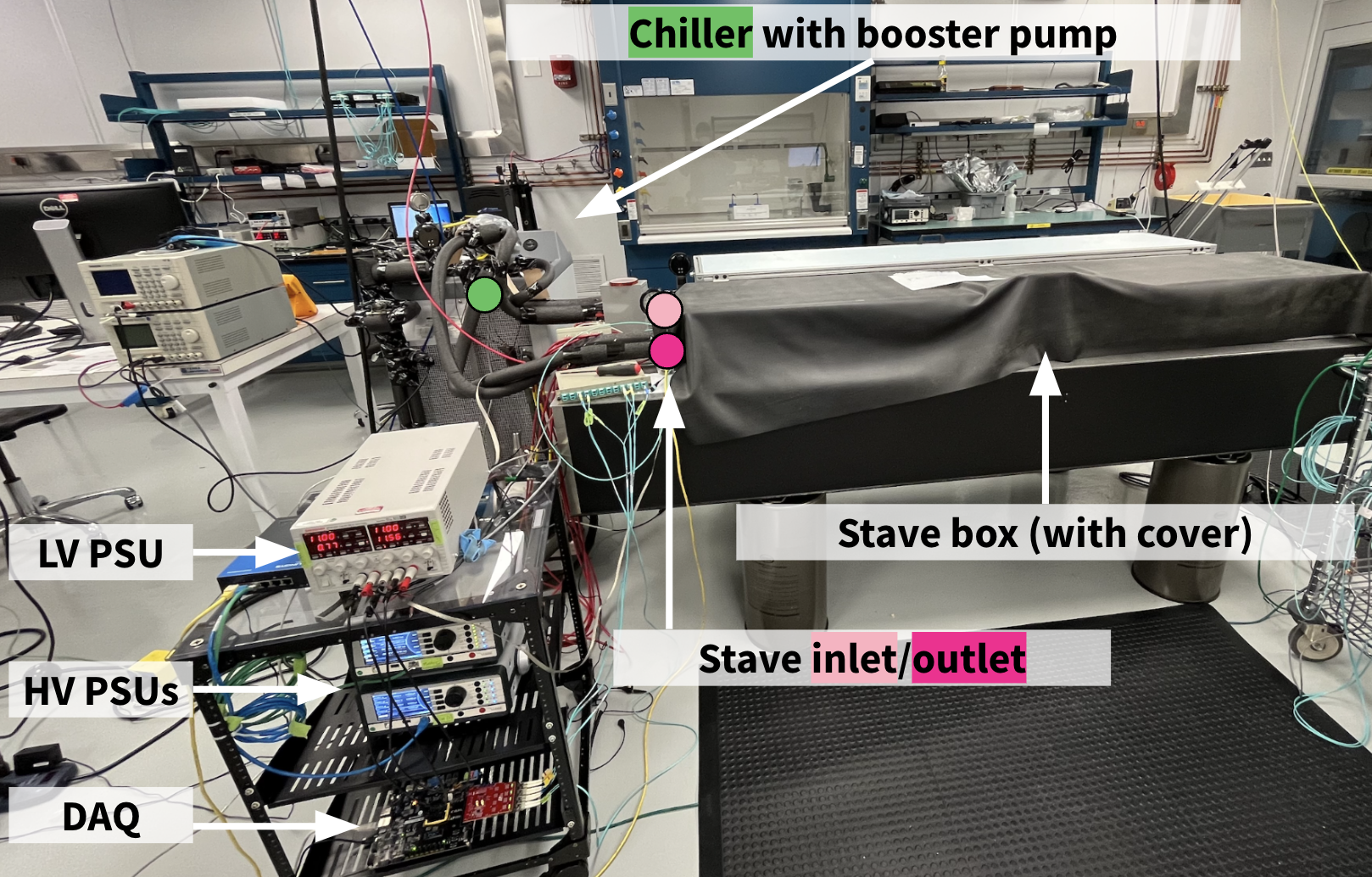}
    \caption{Testing setup at site 2. The following are pictured and labeled: A Keithley low voltage power supply with two channels, one to power the electronics on each side of the stave. Two ISEG SHR high voltage power supplies \cite{ISEG}, one to bias the sensors of each stave side. An SPS chiller with a Lenze ESV751N02YXC NEMA 4x inverter drive \cite{lenze} to control the booster pump, where the booster pump is necessary for increasing the pressure of the SPS chiller coolant for an operational flow rate through the small stave inlet and outlet pipes. A stave box with a cover. When taking electrical IVs, where leakage current is measured as a function of bias voltage, a black tarp is placed on top of the stave box as a precaution to ensure no light leakage when taking cold IVs, necessary due to the light sensitivity of the sensors. A Genesys evaluation board of the Kintex-7 FPGA, attached to a custom readout board, is used for fast data I/O between the testing PC and stave.}
    \label{fig:Testing_Setup_2}
\end{figure}

Site 2 has one stave testing setup, shown in Figure \ref{fig:Testing_Setup_2}.

The connections to the stave are the same as what was shown for site 1 in Figure \ref{fig:Setup_1_StaveTestBox}. As is the case for the first testing setup of Site 1, this stave testing setup is equipped with a software monitoring system, allowing for the constant monitoring and logging of environmental and on-module monitoring quantities, including the relative humidity and temperatures of modules. There are pros and cons of stave testing with monitoring: The major advantage is that having the full information stored with second-level granularity allows one to investigate unexpected sudden changes in behavior for modules and the environment, and to understand if testing procedures ran as expected. The con is that monitoring DAQ quantities requires providing LV to the AMAC, which induces a current of about 50 mA per AMAC, leading to a small but non-negligible heat dissipation on each powerboard. When trying to reach the coldest sensor temperature possible, it is necessary to run with LV off to avoid this heat dissipation, with the trade-off of not having full monitoring. 

Relative humidity, used to compute dew point, is monitored to ensure that there is no risk of condensation when testing at cold temperatures. Various temperatures are also monitored, namely that of the stave inlet and outlet coming from NTC measurements, the temperature of the chiller bath measured internally by the SPS chiller, the set point of the chiller, and the average of the stave inlet and outlet temperatures, used as a benchmark for testing temperature. Novec HFE7100 is used as the cooling fluid, which does not leave residue in the stave cooling pipes. This is a particularly important quality of this choice of cooling fluid, as this ensures that staves which are installed in the detector do not introduce contamination in the final CO$_{2}$ plant which will be used for cooling the detector.

As part of the on-module monitoring quantities, voltage and current of the two HV PSUs is included, for each of the 4 channels on each PSU, where each channel powers a different segment of modules. These are useful for verifying the time periods during which modules were biased, and when IVs were occurring. If the monitoring shows modules are not biased when they are expected to be, it can hint at an issue in the DAQ or software sequence. Additionally, higher than expected HV current values may correspond to higher than expected module leakage current. 

Voltage and current of the two LV PSU channels is also monitored. As with the HV PSUs, monitoring voltage values ensures that the hybrid and powerboard are powered when they should be. Unexpected values can indicate issues in the DAQ or software sequence. Monitoring the current values of the LV PSU channels indicates if the AMAC and hybrids are powered, and these values act as a proxy to estimate the power dissipation on the stave. An example case in which having monitoring was very useful is described in Appendix \ref{appendix:MonitoringExample}.

\subsection{Temperature normalization among setups}

While the expected operating temperature of ITk strips staves is -35$\degree$C, it is possible that the detector may reach lower temperatures. During the operation of the ITk, CO$_{2}$ cooling will be used. While temperatures experienced in non-operational scenarios are difficult to predict, a large leak in the CO$_{2}$ cooling system at atmospheric pressure may cause the coolant to reach temperatures around -55$\degree$C, the freezing point of CO$_{2}$. It is difficult to predict which part of the detector would experience temperatures around this value, and for how long, but it is conceivable for modules to reach this temperature in extreme scenarios. Therefore, it must be ensured that staves can operate at or below this temperature in order to ensure there is headroom with the chosen mitigation strategy. Additionally, in operational conditions, turning off the stave is expected to decrease its temperature by about 5-10 degrees.

\begin{figure}[htbp]
    \centering
    \includegraphics[width=0.9\linewidth]{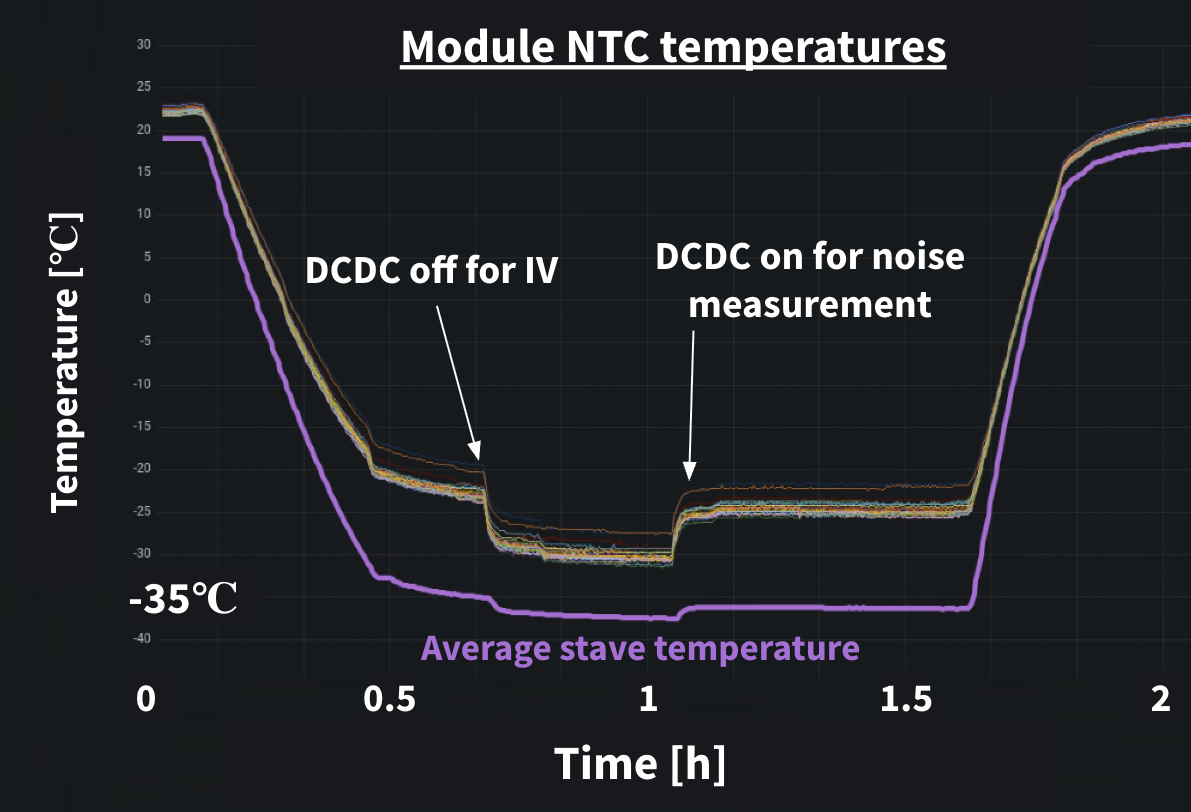}
    \caption{Site 2 average stave and Module NTC temperatures. It can be seen that powering hybrids off or on leads to changes in the average stave and Module NTC temperatures.}
    \label{fig:Setup1_NTCs}
\end{figure}

As stave testing setups are not identical, and it's known that low module temperatures are expected to cause fracturing, it's important to understand the approximate module temperatures and their differences with respect to the stave inlet and output temperatures. One way to achieve this is via the modules' NTC measurements. An example of this is shown in Figure \ref{fig:Setup1_NTCs}. Shown in purple is the average stave temperature, defined as the average of the stave inlet and outlet NTC temperatures, as a function of time during a thermal cycle. The other colors correspond to the 28 NTC measurements, one for each module on a given stave (stave 6 in this case). It can be seen that when the average stave temperature is lowered to approximately -35$\degree$C, the hybrid temperatures are grouped around -24$\degree$C with hybrids powered on. When the hybrids are turned off, the NTC temperatures drop by approximately 5$\degree$C due to the removal of the heat load generated by the current from powering the hybrids. With hybrids on (off), the NTCs are approximately 10-12 (8-10) degrees warmer than the average stave temperature. 

\subsection{Testing procedure}
\label{sec:testingProcedure}

\begin{figure}[htbp]
    \centering
    \includegraphics[width=\linewidth]{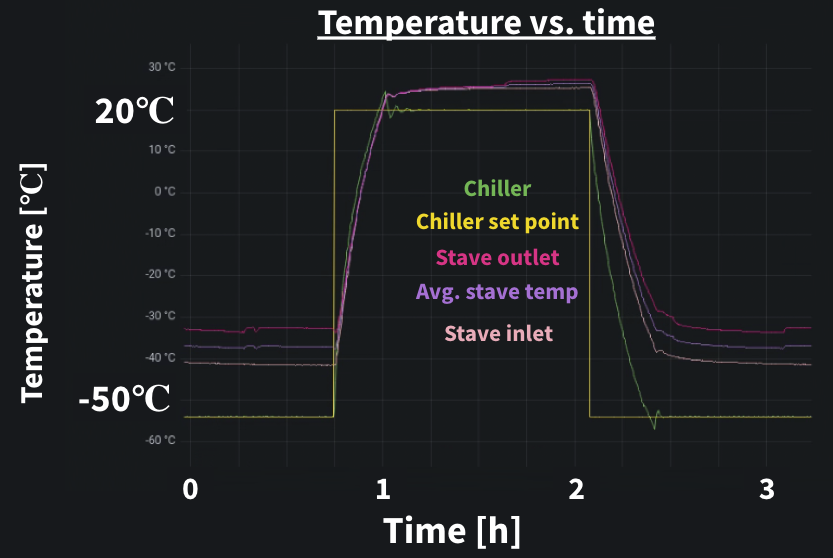}
    \caption{Thermal cycle definition, showing various temperatures of the testing setup as temperature of the coolant is varied.}
    \label{fig:TC_Definition}
\end{figure}

In order to emulate the effect of temperature changes experienced by staves, ``thermal cycling" is performed. As part of the standard quality control procedure, every module must undergo a set of ten thermal cycles before being glued onto a local support structure, as described in \cite{TC}. A similar procedure is performed for modules loaded onto staves, where the definition of a thermal cycle is shown in Figure \ref{fig:TC_Definition}. A thermal cycle is defined as starting at a cold temperature below room temperature, taking an IV measurement of all modules on a stave up to -550V, lowering the bias voltage to -350V in order to take a noise measurement of all modules, increasing to around room temperature, repeating the IV and noise measurement, and cooling back to the desired cold temperature.

Although the normal QC of staves was planned to consist of two cycles down to -35$\degree$C, an extended QC program was performed during the design validation stage when testing staves with different fracturing mitigation strategies. During this design validation stage, staves assembled at Site 1 would typically undergo 5 thermal cycles to an average stave inlet/outlet temperature of about -35$\degree$C, with an IV and noise measurement taken at each cold and warm point. Five thermal cycles to -42$\degree$C are then performed, with an IV and noise measurement again taken at each cold and warm point. Staves at this site are then moved to the climate chamber to undergo the same procedure but at the following temperatures: -50$\degree$C, -55$\degree$C, -60$\degree$C, -65$\degree$C, and finally -70$\degree$C. All tests at temperatures between -35$\degree$C and -55$\degree$C are considered extended-QC, and was only done for these design validation objects. Tests in the climate chamber down to -70$\degree$C are part of the QA (Quality Assurance) only performed during design validation.

\begin{figure}[htbp]
    \centering
    \includegraphics[width=\linewidth]{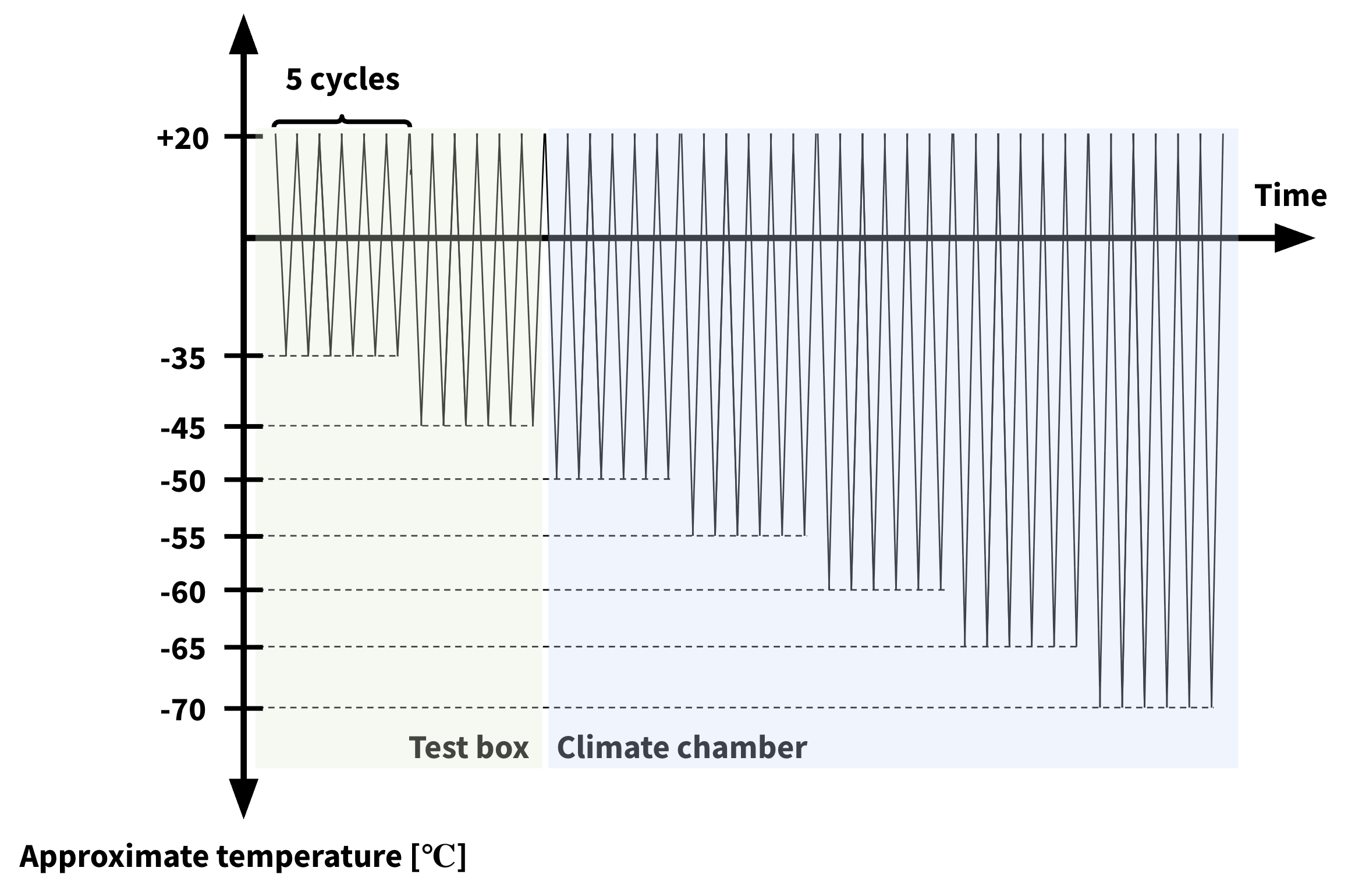}
    \caption{Sketch of the full testing sequence for a stave. 5 thermal cycles are performed down to -35$\degree$C, and -45$\degree$C in stave test boxes. Staves are then transferred to the climate chamber at site 1 where 5 cycles are then performed in steps of -5$\degree$C down to the minimum achievable testing temperature of the climate chamber: -70$\degree$C.}
    \label{fig:full_sequence_sketch}
\end{figure}

For staves assembled at site 2, staves typically undergo 5 thermal cycles to an average stave inlet/outlet temperature of about -35$\degree$C, with an IV and noise measurement taken at each cold and warm point. Five thermal cycles to -45$\degree$C are then performed, with an IV and noise measurement again taken at each cold and warm point. As this is the minimum temperature that site 2 can achieve, staves are then sent to site 1 in order to undergo additional thermal cycling in the climate chamber. These stave typically undergo 5 cycles at each of the following temperatures: -50$\degree$C, -55$\degree$C, -60$\degree$C, -65$\degree$C, and finally -70$\degree$C, the minimum achievable temperature of the climate chamber. A sketch of the full testing procedure is shown in Figure \ref{fig:full_sequence_sketch}. Due to complex testing schedules coordinated among site 1, site 2, and module assembly sites, sometimes staves were not able to undergo this full testing procedure. In these cases, testing at initial temperatures was expedited in favor of testing at -70$\degree$C. 

Testing at increments of -5$\degree$C is desirable when testing mitigation staves, and was performed in order to identify the temperatures at which each design begins to fail, to understand the headroom of each design separately. This level of granularity was not necessary between -35$\degree$C and -45$\degree$C, as this is already very close to the operational temperature of the ITk. 

\section{Results}
\label{sec:Results}

The 8 staves assembled at site 1 and site 2, as previously summarized in Table \ref{fig:Stave_Summary}, were subject to extended QC in order to evaluate the different mitigation strategies. This section will describe the results of modules on these staves, with the following temperatures and conditions: testing of modules down to -70$\degree$C with high statistics, and an extreme QA test where an interposer stave was cycled 100 times to -45$\degree$C. 

\subsection{Temperature varied testing}

\begin{figure}[htbp]
    \centering
    \includegraphics[width=\linewidth]{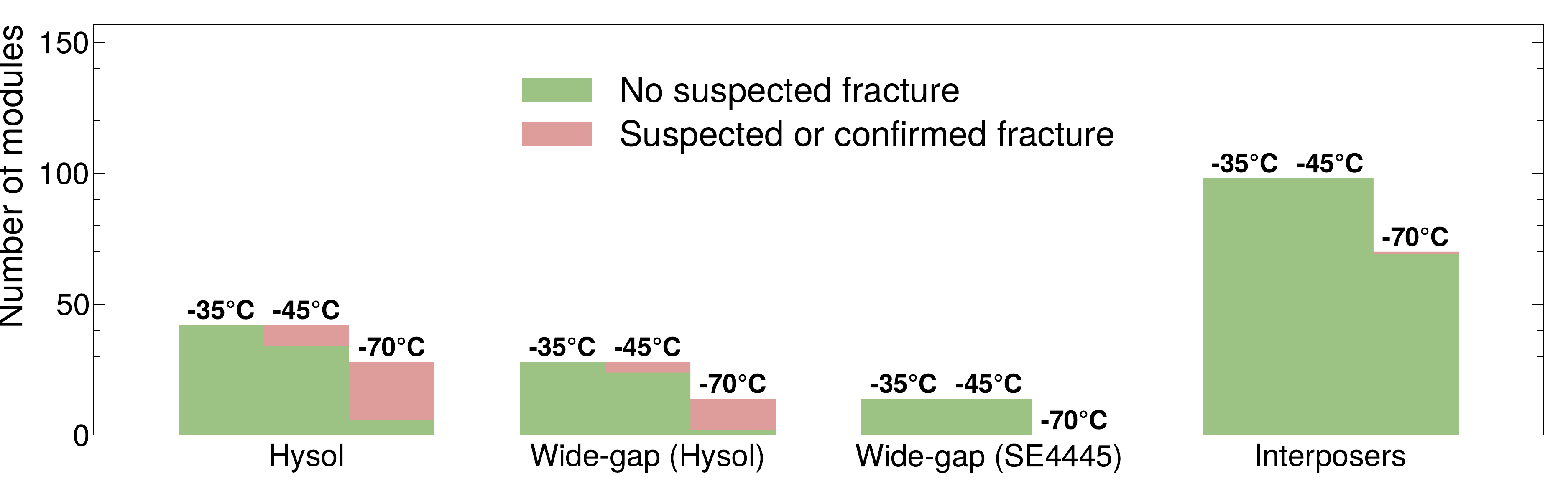}
    \caption{Summary of results for different mitigation strategies, showing the number of modules tested at each temperature. ``No suspected fracture" corresponds to modules which do not exhibit early breakdown from IV tests, nor sporadic noise measurements. ``Suspected or confirmed fracture" corresponds to modules which show early breakdown or sporadic noise measurements.}
    \label{fig:Results_Summary}
\end{figure}

A summary of all modules tested at the two stave testing setups is shown in Figure \ref{fig:Results_Summary}. This shows that while hysol loaded modules do not fail at -35$\degree$C, they begin to fail at -45$\degree$C, and fail at a higher rate at -70$\degree$C. This is consistent with the understanding of the sensor fracturing mechanism, through which more strain is expected from colder temperatures.

Wide-gap modules loaded with hysol show a similar trend: they do not fracture at -35$\degree$C, they start to fracture at -45$\degree$C, and at a higher rate at -70$\degree$C. This shows that hysol is not a viable solution to mitigate sensor fracturing. It is noted that no wide-gap modules loaded with SE4445 are suspected to have fractured, but these were not tested to -70$\degree$C as wide-gap modules were already observed to fracture with Hysol, and are a non-ideal mitigation strategy as they would not work for the SS module design.

Finally, it can be seen that no interposer modules tested down to -45$\degree$C have suspected fractures, and only one module out of 70 which were tested down to -70$\degree$C are suspected to have fractured. The lowest temperature at which this module was still functional was a -65$\degree$C air temperature.

\begin{figure}[htbp]
    \centering
    \includegraphics[width=\linewidth]{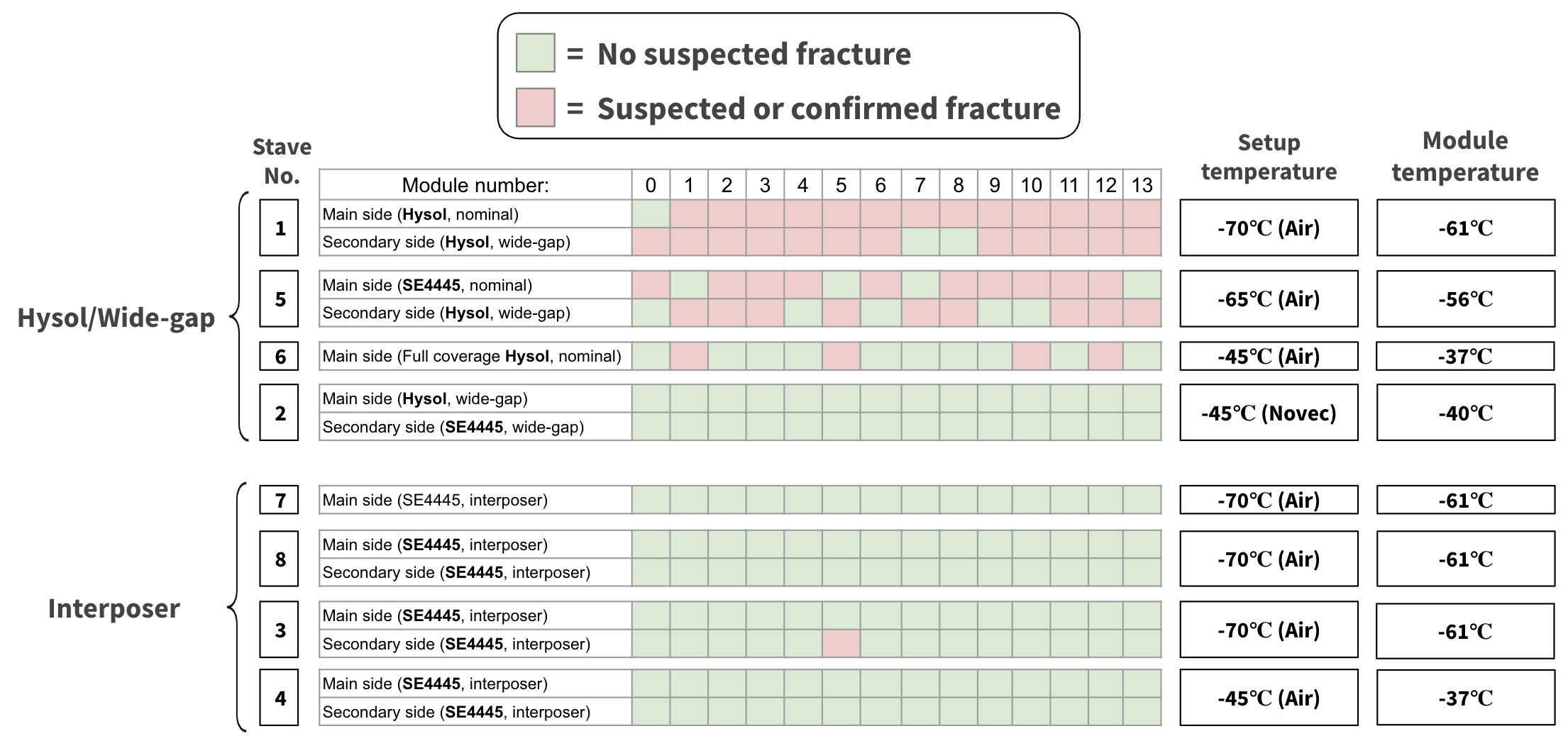}
    \caption{Per-module results for the 8 mitigation staves, separated by mitigation strategy, with setup and effective module temperature specified.}
    \label{fig:PerModule_Results}
\end{figure}

A summary of results per module is shown in Figure \ref{fig:PerModule_Results}. The same trend can be seen from the Hysol/Wide-gap modules, in that most of them fail by the time they reach -70$\degree$C. From the interposer module results, it is seen that almost all modules tested down to -70$\degree$C are not suspected to have fractured. It was also initially seen that three modules show an early breakdown. After deionizing these modules and drying out the stave for two weeks, these breakdowns went away.

\begin{figure}[htbp]
    \centering
    \includegraphics[width=\linewidth]{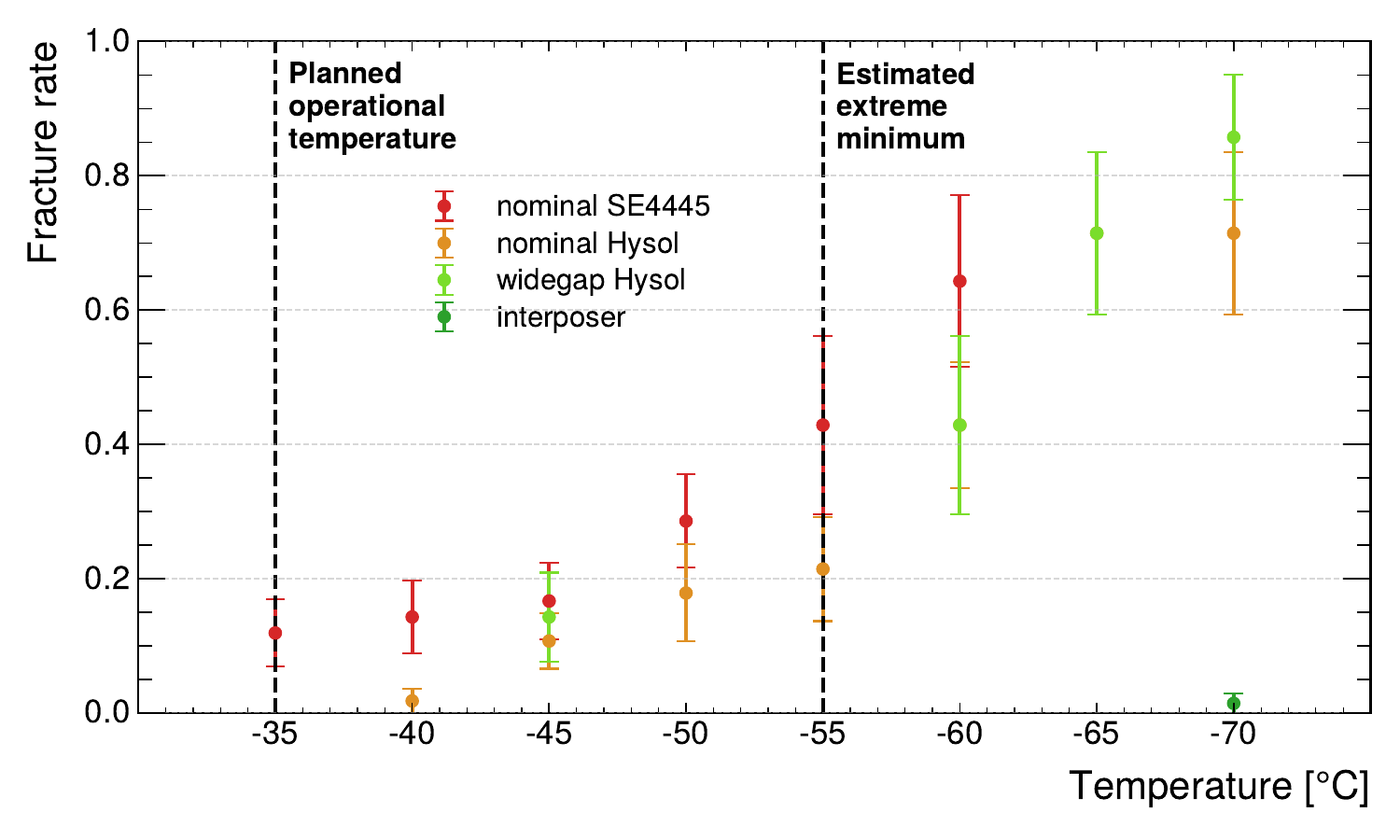}
    \caption{Fracture rate as a function of temperature for the nominal module design, and mitigation designs.}
    \label{fig:VsTempResults}
\end{figure}

Finally, the results are shown for each mitigation strategy and the original nominal SE4445 design as a function of temperature in Figure \ref{fig:VsTempResults}. This shows that nominal SE4445 modules begin fracturing at the ITk operational temperature, replacing SE4445 with Hysol lowers the initial temperature at which modules begin fracturing, widening the gap between the Hybrid and powerboard further lower this initial fracturing temperature, and using interposers dramatically decreases this temperature.

These results indicate that the interposer design is the best performing solution tested here to mitigate sensor fracturing in the barrel, as no modules are seen to fracture at or slightly past the expected operational temperature of -35$\degree$C, and only one out of 70 modules is seen to fracture at -70$\degree$C, much colder than the expected operating temperature.

\subsection{Quality assurance}

All staves must pass the extended-QC procedure (tests performed between -35$\degree$C to -55$\degree$C) and QA (tests performed below -55$\degree$C) discussed above, but it is important to confirm that sensors do not fracture in the cycles immediately beyond the standard amount. To ensure the robustness of interposer modules on staves, Stave 8 was thermally cycled between $\approx$~20$\degree$C and -45$\degree$C 100 times. In a previous QA study, four ITk Strips barrel modules were thermal cycled 100 times, but these modules were off-stave \cite{TC}. This quality assurance cycling of Stave 8 was conducted at Site 2.

At the start, stave 8 was cycled identically to nominal stave cycling. During the first 5 cycles, the standard suite of tests was performed at each cold and warm point. As cycles continued, testing was conducted less frequently, as the main purpose of the test was to check the final results after all cycles rather than the progression of results as a function of cycle number. Roughly, testing slowed from a frequency of once per cycle, to once every 5 cycles, to once every 10 cycles, and eventually to once every 20 cycles. The focus of this robustness testing is endurance, not precision. When no testing was involved, cycling was conducted with the LV off. After cooling, the stave was held at an average stave inlet/outlet temperature of -45$\degree$C for 5 minutes to allow components to equilibrate. 

After 26 thermal cycles, 5 of the modules on Stave 8 experienced electrical damage which prevented them from being tested further. It is believed that one of the HV PSUs erroneously powered on two HV channels and output $\approx$~2000~V for a period of several hours overnight. For an unknown reason, this was not registered by the monitoring system. While this event was not explicitly confirmed, in a future instance this type of behavior was observed on the physical PSU monitor, motivating the hypothesis. Regardless, as a result, the 6 modules on these HV channels are excluded from the QA dataset.

\begin{figure}[!htb]
    \setcounter{subfigure}{0}
    \centering
    \subfloat[Stream 0, corresponding to strips underneath the hybrid and powerboard. \label{fig:stream0_RT}]{\includegraphics[width=0.485\textwidth]{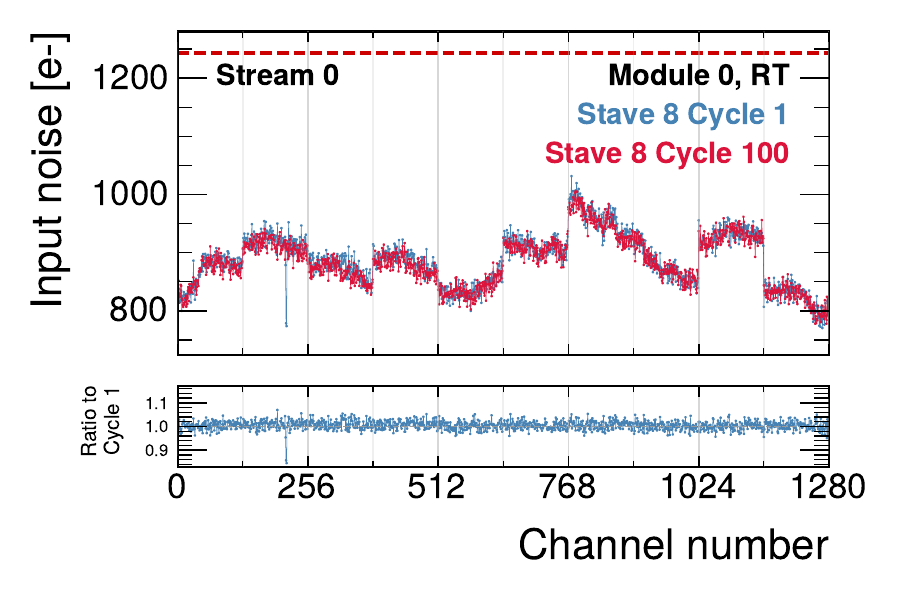}}%
    \hfill
    \subfloat[Stream 1, corresponding to strips uncovered by the hybrid and powerboard. \label{fig:stream1_RT}]{\includegraphics[width=0.485\textwidth]{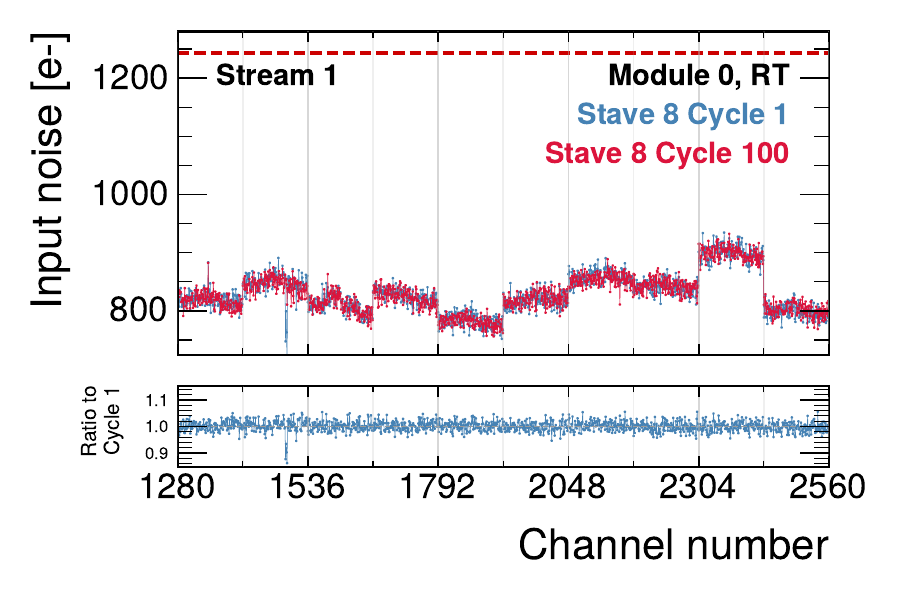}}%
    \caption{\label{fig:1000noise_RT}Comparison of module noise for an example module on Stave 8 in the initial and final room temperature tests. The dashed red line indicates the maximum allowed noise of 1243 ENC.}
\end{figure}

\begin{figure}[!htb]
    \setcounter{subfigure}{0}
    \centering
    \subfloat[Stream 0, corresponding to strips underneath the hybrid and powerboard. \label{fig:stream0_Cold}]{\includegraphics[width=0.485\textwidth]{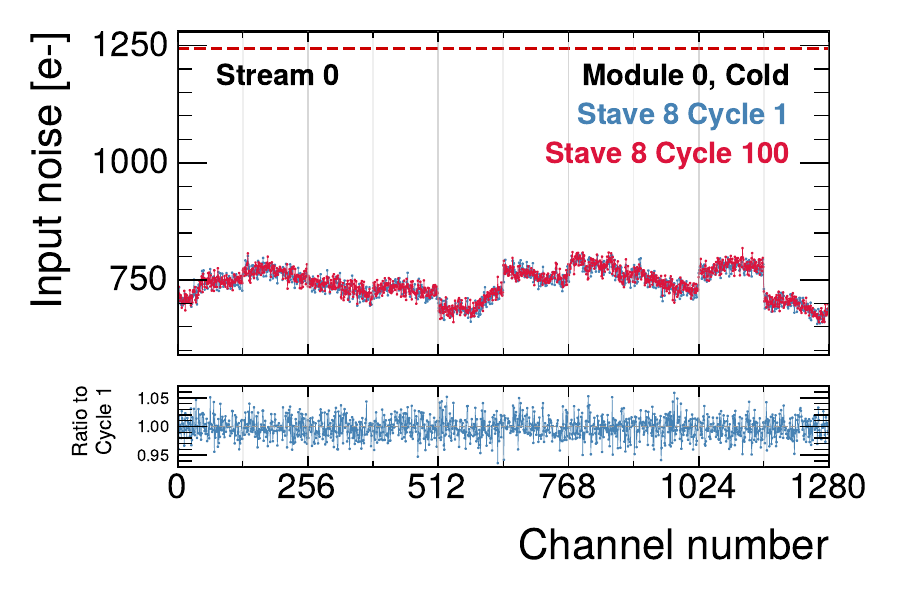}}%
    \hfill
    \subfloat[Stream 1, corresponding to strips uncovered by the hybrid and powerboard. \label{fig:stream1_Cold}]{\includegraphics[width=0.485\textwidth]{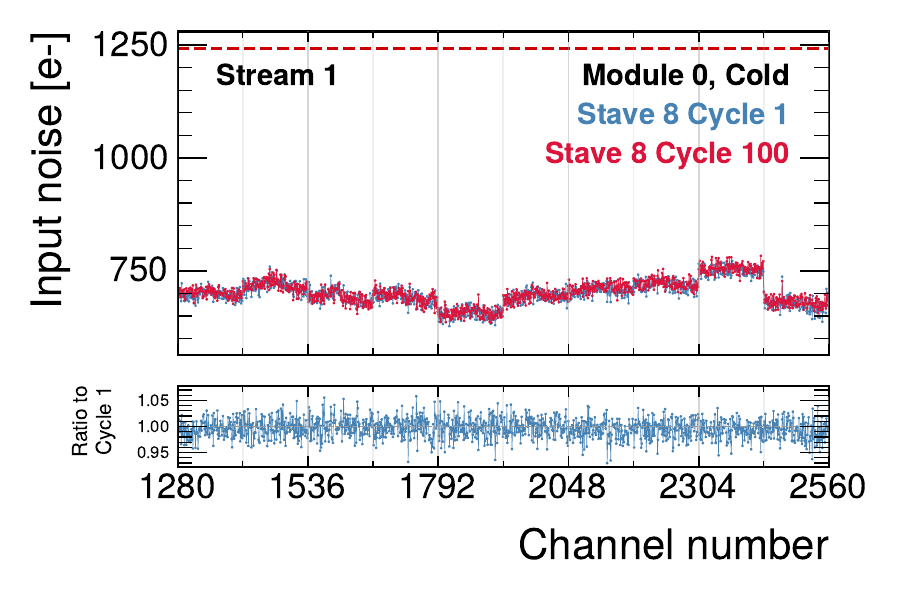}}%
    \caption{\label{fig:100noise_cold}Comparison of module noise for an example module on Stave 8 in the initial and final cold tests at -45$\degree$C. The dashed red line indicates the maximum allowed noise of 1243 ENC.}
\end{figure}

\begin{figure}[!htb]
    \setcounter{subfigure}{0}
    \centering
    \subfloat[Room temperature tests. \label{fig:IV_RT}]{\includegraphics[width=0.485\textwidth]{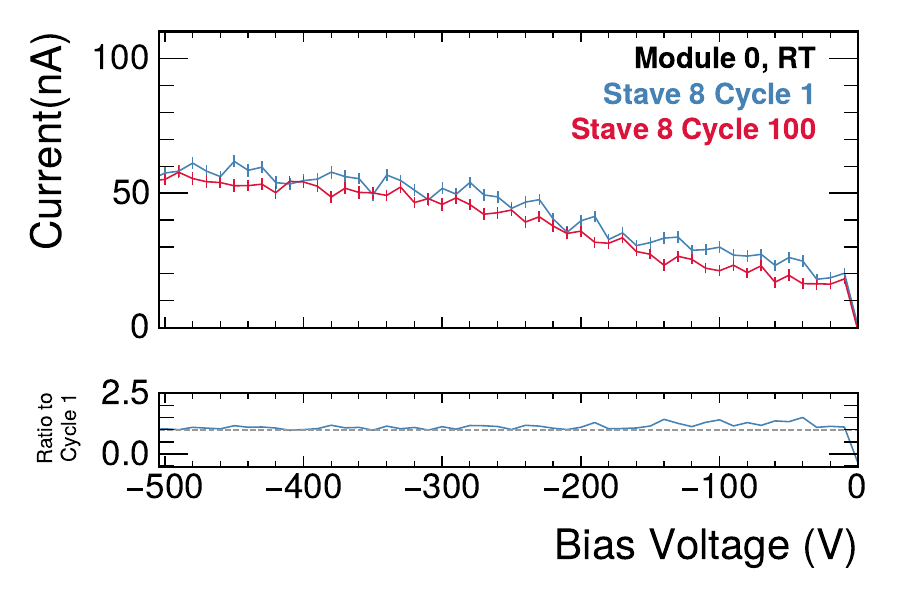}}%
    \hfill
    \subfloat[Cold tests. \label{fig:IV_Cold}]{\includegraphics[width=0.485\textwidth]{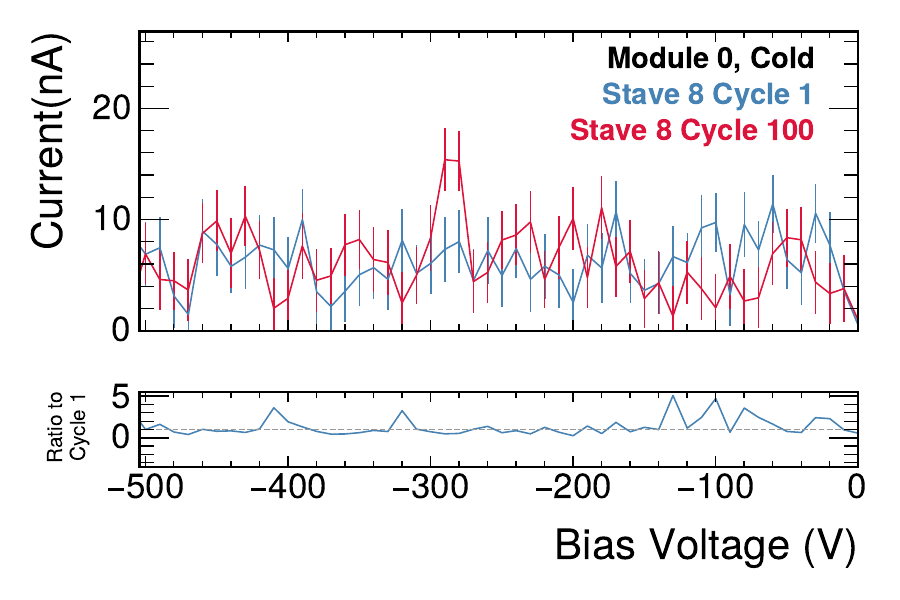}}%
    \caption{\label{fig:100IV}Comparison of IV scans for an example module on Stave 8 in the initial and final cycle.}
\end{figure}

After 100 cycles, none of the remaining 22 modules showed any signs of early breakdown or fracturing. Comparisons of the initial and final noise measurements for an example module are shown in Figures \ref{fig:1000noise_RT} and \ref{fig:100noise_cold}, with these figures representing warm tests and cold tests respectively. Similarly, a comparison of initial and final IV scans for an example module is shown in Figure \ref{fig:100IV}. These results are similar to what is seen for all modules - very similar noise and IV measurements after 100 thermal cycles, indicating interposer modules on staves are robust to a high number of thermal cycles.

Additional QA was performed using Stave 3 at Site 1 to further test the limits of interposer stave sides. When testing this stave in the climate chamber to -70$\degree$C at Site 1, the stave was left unpowered in order to reach as low of a temperature as possible. This stave was also warmed to about 30$\degree$C, instead of 20$\degree$C, in order to test a more extreme warm temperature.
\section{Conclusions}
\label{sec:conclusions}

As a result of the discovery of sensor fracturing of local-support-loaded ITk strips barrel modules, multiple mitigation strategies were tested. These mitigation strategies were tested at the two local support loading sites for the barrel, which are able to test modules at different temperatures and with different cooling methods. Staves built at these two sites yield compatible testing results. These results show that the most performant strategy is the interposer design, due to the extremely low rate of problematic modules at temperatures lower than expected during operations - only 1 out of 70 modules exhibited behavior consistent with module fracturing at -70$\degree$C, indicating this design should be robust at the ITk operational temperature of -35$\degree$C. This is bolstered by the QA test performed on interposer modules on staves, which showed very similar noise and IV characteristics after 100 thermal cycles to -45$\degree$C.

The design verification stage is almost complete. Further staves are currently being loaded and tested with interposer modules in order to increase the size of the interposer stave dataset. Additionally, radiation tests are being performed. A total of 8 additional staves will undergo extended-QC to ensure a large statistical dataset of interposer staves with an acceptable yield, before moving to the full production of LS staves for the ITk strips barrel. Additional design verification tests will be performed for SS staves with interposers.


\appendix

\section{Assembly details}
\label{appendix:AssemblyDetails}

\begin{figure}[htbp]
    \centering
    \includegraphics[width=\linewidth]{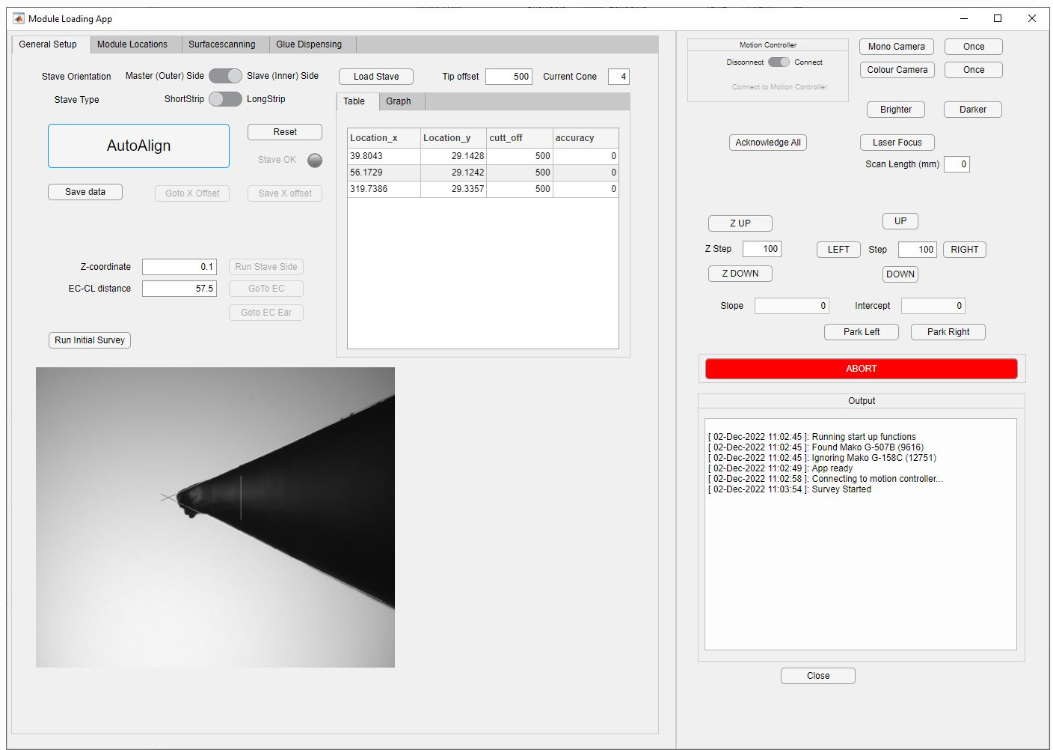}
    \caption{A screenshot of the loading software used at site 1, during the calibration of the gantry. The cones on the core's lockpoints are found to determine the stave co-ordinate x-axis.}
    \label{fig:Site_1_Lockpoint}
\end{figure}

\begin{figure}[!htb]
    \centering
    \includegraphics[width=0.5\linewidth]{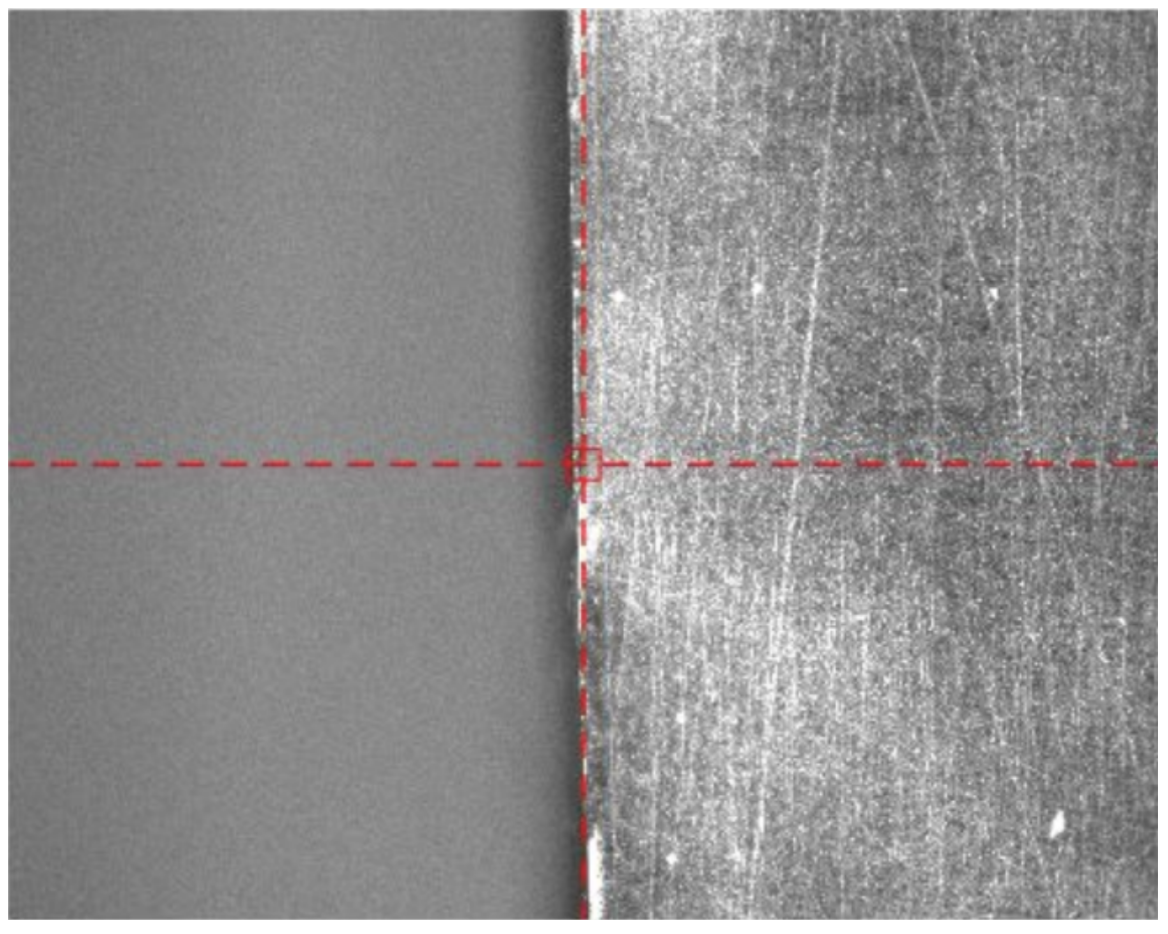}
    \caption{A screenshot of the loading software at site 1, during the calibration of the gantry. The z=0 edge of the core in the middle of the core's surface is located.}
    \label{fig:Site_1_CoreEdge}
\end{figure}

Before any actions are performed using the gantry, a calibration is performed to find the stave core on the table and convert gantry coordinates to stave coordinates. This is done by finding lockpoints on the core, as they are machined with high precision (micron level) and serve as a reference point. The loading software first uses the camera to find the cones on the core's lockpoints, where pattern recognition is used to identify the tips of the cones. A straight line is fit through the cones to determine the stave co-ordinate x-axis, an example of which is shown in Figure \ref{fig:Site_1_Lockpoint}. The loading software then finds the z=0 edge of the core in the middle of its surface, and together with the previous straight line, defines a co-ordinate system on the stave core that aligns the axes directions with the z-axis expected at the ATLAS detector. An example of the edge finding software is shown in Figure \ref{fig:Site_1_CoreEdge}.

\begin{figure}[!htb]
    \centering
    \includegraphics[width=0.5\linewidth]{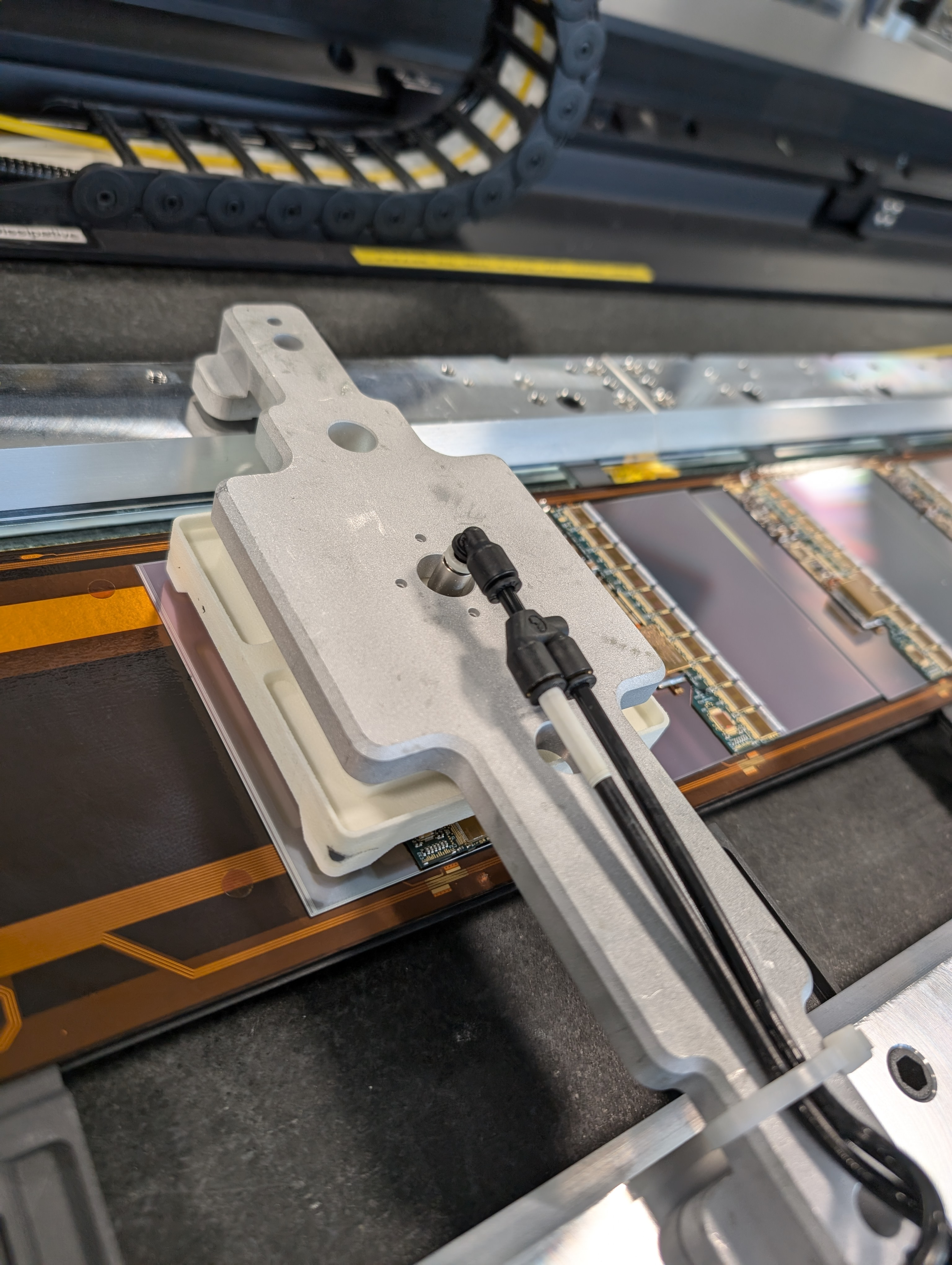}
    \caption{The module pickup tool, which uses vacuum at the four corners of the module. The tools have dowel pins that sit in holes in the assembly frame giving an initial gross alignment of the module.}
    \label{fig:Site_1_ModulePickUpTool}
\end{figure}

The module pickup tool possesses dowel pins that slot into holes in the assembly frame, giving a gross location of where the module should be positioned, as shown in Figure \ref{fig:Site_1_ModulePickUpTool}. The vacuum is then released and the tool is removed, leaving the module sitting on the glue. Similar to site 2, kapton spacers are used at each of the four corners of the module to set the glue height. The module's stereo fiducial markers are then located using the gantry camera and the module loading software, and fine adjustments are made to the module's position by gently nudging it with an ESD safe stick whilst monitoring the fiducials on a display screen.

\section{Monitoring example}
\label{appendix:MonitoringExample}

\begin{figure}[htbp]
    \centering
    \includegraphics[width=0.9\linewidth]{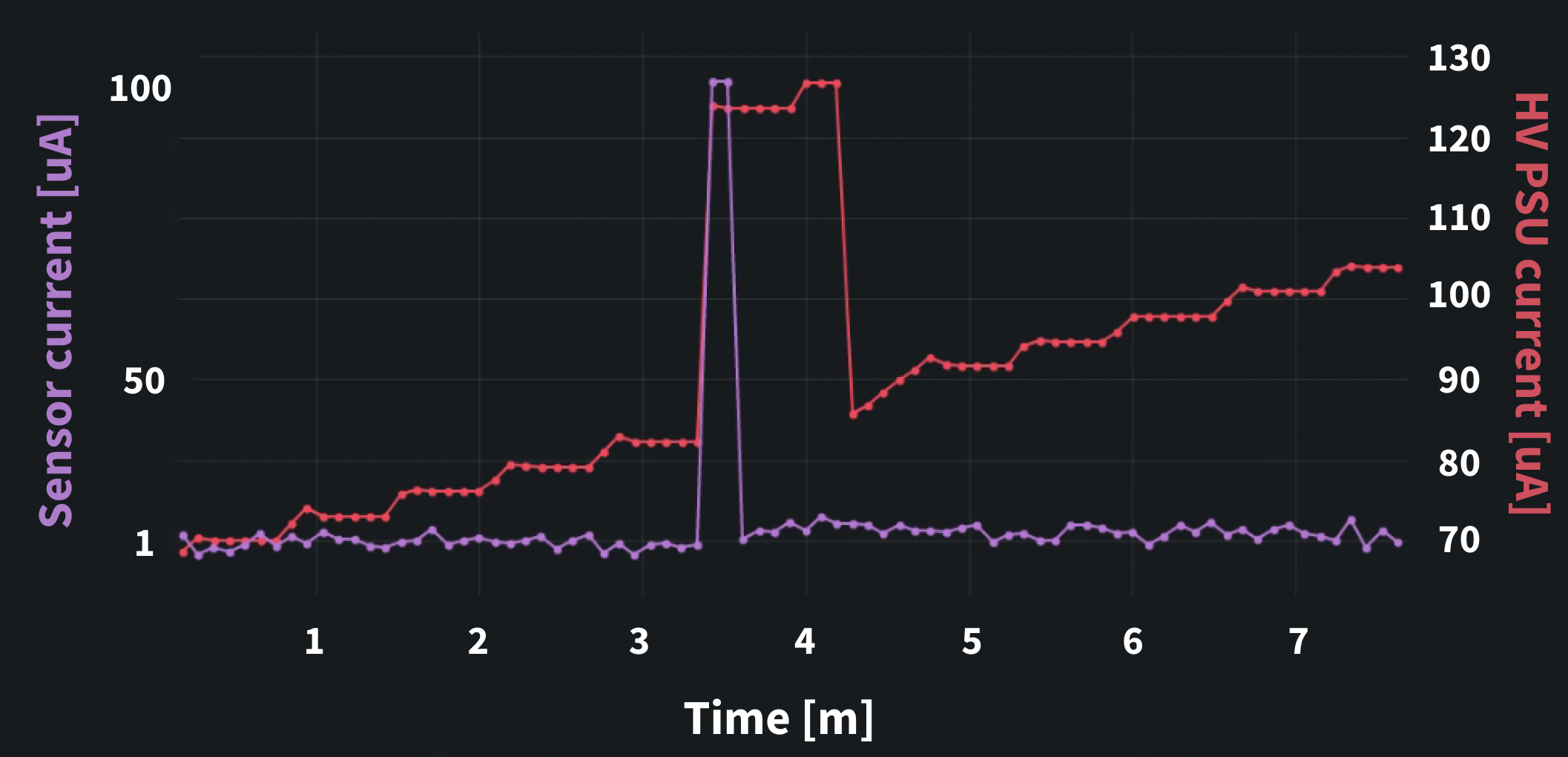}
    \caption{Suspected moment of module fracture.}
    \label{fig:suspected_fracture_moment}
\end{figure}

\begin{figure}[htbp]
    \centering
    \includegraphics[width=0.9\linewidth]{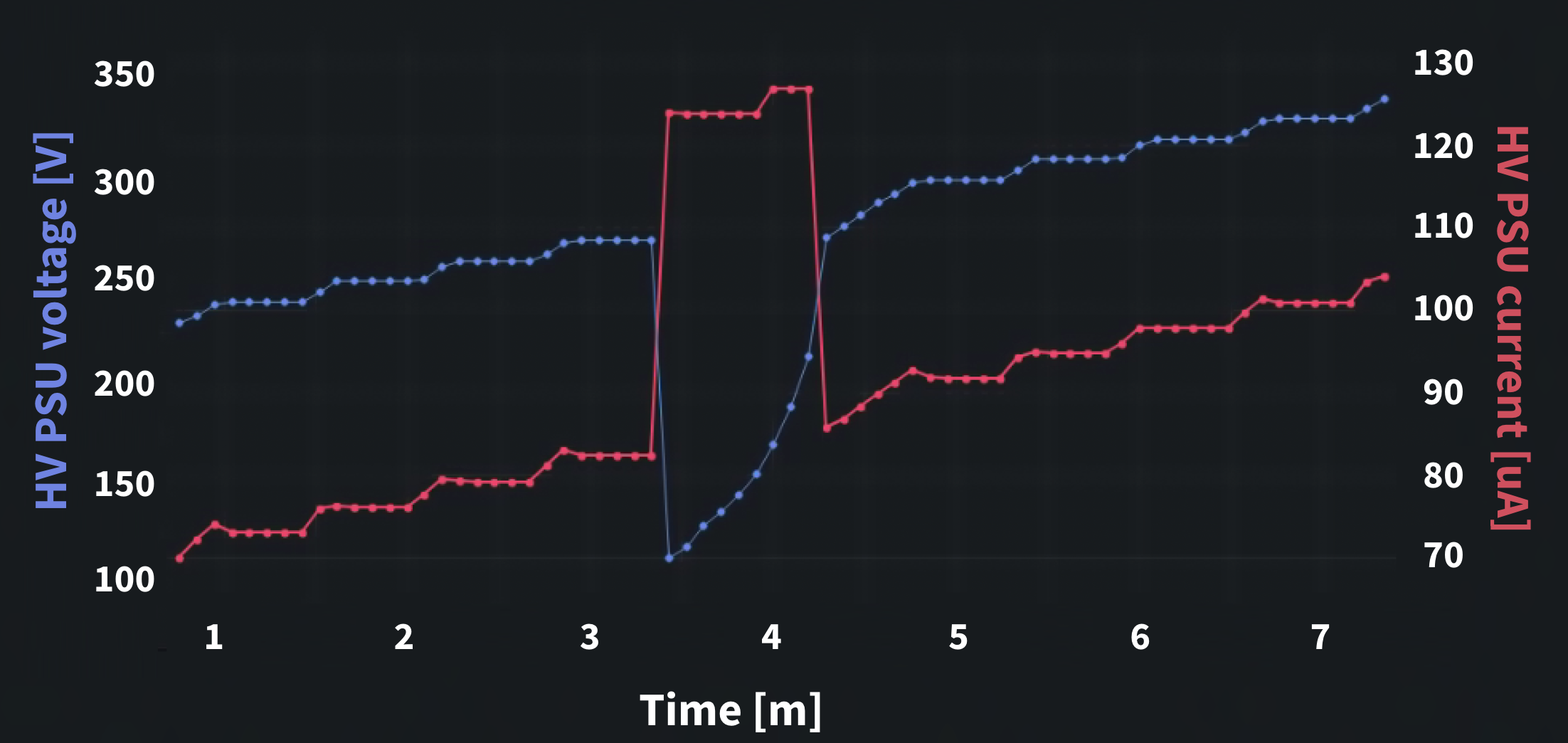}
    \caption{HV PSU voltage and current during suspected fracture event.}
    \label{fig:HV_PSU_fractureMoment}
\end{figure}

During the testing of Stave 5, during its first set of cold tests, a spike was observed in a module's sensor current, shown in Figure \ref{fig:suspected_fracture_moment} in purple. The sensor current reading is reported to be approximately 100~$\upmu$A, which would be much higher than expected, as a power supply current of approximately 35~$\upmu$A is expected when the sensor is biased at -350V. This expected power supply current is dominated by a 10 M$\ohm$ resistor in the powerboard, in parallel with the sensor in the circuit. This sensor current reading of 100~$\upmu$A is not expected to be exact as the AMACs ADC circuits were saturated, meaning it's not possible for the sensor measurement circuit to report anything larger. However, the power supply current measurement from this module's corresponding HV segment shows a change at the same time, shown in Figure \ref{fig:HV_PSU_fractureMoment} in red, where an increase of approximately 40~$\upmu$A is seen. The PSU was in constant current mode \cite{ISEG_Manual}, meaning that if the maximum allowed current is reached, the PSU will change its voltage in order to try and keep this current constant. This is one of multiple modes of the power supply. Following this protocol, at this moment, the voltage of the PSU lowered, and eventually returned to nominal values around minute 4.5, as shown in Figure \ref{fig:HV_PSU_fractureMoment}. This indicates the module saw a current of approximately 40~$\upmu$A for at least a moment, but that current returned to nominal values. This would be a large current value for a module to see at this voltage, as at 80 volts, the HV PSU is expected to read approximately 8~$\upmu$A for a module. For typical healthy LS modules, module leakage currents on the order of 100 nA are seen up to bias voltages of -550V. 

During the next room temperature electrical test of this module, the leakage current increased to the order of $\upmu$A before reaching -200V, considered an ``early breakdown", indicating the sensor may have been fractured. It is possible for problematic sensors to reach leakage current magnitudes consistent with breakdown closer to -550V, but these breakdowns are not as consistent with fractured sensors. While a fracture was never visually observed for this module as it was expected to be under the hybrid or powerboard, its subsequent electrical tests were consistent with a fractured module, and the monitoring indicates the suspected moment at which the sensor fractured, and can hint at the amount of current generated from the event and the duration for which it propagated through the module and HV PSU.

\end{document}